\documentstyle[12pt]{article}
\input{epsf}
\def\sss{\scriptscriptstyle}
\def\barp{{\raise.35ex\hbox{${\sss (}$}}---{\raise.35ex\hbox{${\sss
)}$}}}
\def\bdbarp{\hbox{$B_d$\kern-1.4em\raise1.4ex\hbox{\barp}}}
\def\bsbarp{\hbox{$B_s$\kern-1.4em\raise1.4ex\hbox{\barp}}}

\newcommand{\beq}{\begin{equation}}
\newcommand{\eeq}{\end{equation}}
\newcommand{\absvcb}{\vert V_{cb}\vert}
\newcommand{\absvub}{\vert V_{ub}\vert}

\newcommand{\absvts}{\vert V_{ts}\vert}

\def\mt{m_t}
\def\mb{m_b}

\def\ml{m_\ell}
\def\Mw{M_W}

\newcommand{\bgamaxs}{$B \to X _{s} \gamma$}

\newcommand{\BBGAMAXS}{{\cal B} (B \to X _{s} \gamma)}
\newcommand{\brbgamaxs}{${\cal B} (B \to X _{s} \gamma)$}
\newcommand{\brbksgam}{${\cal B}(B \to K^* \gamma)$}
\newcommand{\bksgam}{ $B \to K^* \gamma$}
\newcommand{\brogam}{ $B \to \rho \gamma$}

\newcommand{\bomegam}{ $B \to \omega  \gamma$}

\newcommand{\bsgam}{ $b \to s \gamma$}
\newcommand{\bsggam}{ $b \to s \gamma + g$}

\newcommand{\Bsell}
   {$B \to X_s ~\ell^+ \ell^-$}
\newcommand{\Bdell}
   {$B \to X_d ~\ell^+ \ell^-$}

\newcommand{\bksell}
   {$B \to K^* ~\ell^+ \ell^-$}

\newcommand{\bbar}{$B^0$--${\overline{B^0}}$}

\newcommand{\as}{\mbox{$\alpha_{\displaystyle  s}$}}

\def\qbar{\overline q}
\def\sLbar{\overline s_L}
\def\sbar{\overline s}

\def\q5q{\qbar{{\lambda_a}\over 2} i\gamma_5 q}

\def\ebar{\overline l}
\def\gmu{\gamma_\mu}
\def\gmuu{\gamma^\mu}
\def\to{\rightarrow}

\def\mb{m_b}

\def\sig{\sigma_{\mu \nu}}

\def\VV{\ebar \gmuu l}
\def\AA{\ebar \gmuu \gamma_5 l}

\def\s{\hat s}

\begin{document}
\begin{flushright}
CERN-TH.7346/94\\
\end{flushright}
\begin{center}
{\Large \bf
\centerline
{Towards a Model-Independent Analysis of Rare $B$ Decays}}
\vspace*{1.5cm}
\vskip1cm
 {\large A.~Ali}\footnote{On leave of absence from DESY, Hamburg, FRG.},
 {\large G.F.~Giudice}\footnote{On leave of absence from INFN, 
                                Sezione di Padova, Italy.},
 {\large and T.~Mannel}
\vskip0.2cm
       Theory Division, CERN  \\
       CH-1211 Geneva 23, Switzerland \\
\vskip1cm
{\Large Abstract}
\vspace*{5mm} \\
\parbox[t]{\textwidth}{
Motivated by the experimental accessibility of rare $B$ decays in
the ongoing and planned experiments, we propose to undertake
a model-independent analysis of the inclusive decay rates and
distributions in the processes \bgamaxs~ and  \Bsell ~($B=B^\pm$ or
$B^0_d$). We show how measurements
of the decay rates and distributions in these processes
would allow us to extract the magnitude and sign of the dominant
Wilson coefficients of the magnetic moment operator
$\mb \bar{s}_L \sigma_{\mu \nu} b_R F^{\mu \nu }$ and the four-fermion
operators $(\bar{s}_L \gamma_\mu b_L)(\bar{\ell} \gamma^{\mu} \ell)$
and $(\bar{s}_L \gamma_\mu b_L)(\bar{\ell} \gamma^{\mu}\gamma^5 \ell)$.
Non-standard-model effects could thus manifest themselves at low energy
in rare $B$ decays through the Wilson coefficient having values
distinctly different from their standard-model counterparts.
We illustrate this possibility using 
the examples of the two-doublet Higgs models and the
minimal supersymmetric models. The dilepton invariant mass
spectrum and the forward-backward asymmetry of $\ell^+$ in the 
centre-of-mass system of the dilepton pair in the decay \Bsell ~are
also worked out for the standard model and some representative 
solutions for the other two models. }
\end{center}
\noindent
CERN-TH.7346/94\\
July 1994
\newpage
\textheight 23.0 true cm

\section{ Introduction}
The measurement of the decay mode \bksgam ~by the CLEO collaboration
\cite{CLEOrare1}, having a branching ratio
\brbksgam ~$=(4.5 \pm 1.0 \pm 0.9)\times 10^{-5}$, has put the physics
of the electromagnetic penguins on an experimental footing.This
measurement and the experimental upper bound on the inclusive decay
\brbgamaxs ~$< 5.4 \times 10^{-4}$ at 90\% C.L. \cite{CLEOrare2}
have been analysed in the context of the standard model (SM) \cite{ag5}
and in extensions of it such as the two-Higgs-doublet
models (2HDM) \cite{bsg,hmt,bur}, the minimal supersymmetric models
(MSSM) \cite{bbmr,bg,susy}, and a number of other more exotic
variations \cite{Jhewett,Rizzo94}.  Very recently, 
the CLEO collaboration has published a first measurement of the inclusive 
decay rate, based on the measurement of the photon energy spectrum in 
the decay $B \to X_s \gamma$ \cite{CLEObsg}. The branching ratio 
\begin{equation} \label{incl}
{\cal B} (B \to X_s \gamma ) = (2.32 \pm 0.51 \pm 0.32 \pm 0.20) \times 10^{-4}
\end{equation}
puts more restrictive bounds on the non-SM parameters.
In the SM context, the short-distance
contributions in these transitions are dominated by the top quark and
hence they provide valuable information about its mass and
the Cabibbo-Kobayashi-Maskawa (CKM) weak mixing matrix elements
$V_{ts}V_{tb}$ \cite{CKM}. Specifically, the recent CLEO measurement of
\brbgamaxs ~yields at present the following bounds \cite{ag5}:
\beq
0.62 \leq \left\vert {V_{ts} \over V_{cb}} \right\vert \leq 1.1~.
\label{vtslim}
\eeq
Alternatively, the ratio $V_{ts} / V_{cb}$ can be determined from unitarity 
and one may use $\absvts/\absvcb \simeq 1$ to
obtain from the CLEO measurements bounds on the
Wilson coefficient $C_7(\mb)$ of the effective magnetic moment operator. 
Using ${\cal B} (B \to X_s \gamma ) = (2.32 \pm 0.67) \times 10^{-4}$, 
obtained by adding the statistical and the systematic errors in quadrature, 
one obtains 
\beq
0.22 \leq \vert C_7(\mb) \vert \leq 0.30.
\label{c7bound}
\eeq
This bound is subject to the residual next-to-leading order corrections. 
The inclusive branching ratio ${\cal B} (B \to X_s \gamma ) = 
(2.32 \pm 0.67) \times 10^{-4}$ is consistent with the exclusive
branching ratio ${\cal B}(B \to K^* + \gamma) = (4.5 \pm 1.0 \pm 0.9) 
\times 10^{-5}$, with $R(K^*/X_s) \equiv \Gamma (B \to K^* + \gamma)
/\Gamma (B \to X_s + \gamma )$ calculated to be 
$ 0.1 \leq R(K^*/X_s) \leq 0.2 $ in most theoretical models of recent 
vintage \cite{ag5}, \cite{BKST1}-\cite{Apebkstar}. 

The bound on $\vert C_7(\mb)\vert$ given in (\ref{c7bound}) can be
used to constrain the non-SM contribution to the decay rate \brbgamaxs .
The inclusive branching ratio
in the SM, including the leading corrections in the anomalous dimension
matrix and $O(\alpha_s)$ virtual and bremstrahlung corrections,
has been estimated to be
\brbgamaxs $= (2.8 \pm 1.0) \times 10^{-4}$ \cite{ag5}. With
$\mt = 174 \pm 16$~GeV determined from the CDF data \cite{CDFmt},
a good part of the uncertainty in the theoretical
estimates of \brbgamaxs ~is due to the scale dependence of the
perturbative QCD framework and the QCD scale parameter entering
\as ~itself. Recently, parts of the next-to-leading-order
corrections have been
included in the anomalous dimension matrix and the resulting
decay rate $\Gamma (B \to X_s + \gamma)$ has been recalculated
\cite{Apebkstar}. The inclusion of these terms
reduces the scale-dependence of the effective
coefficient $C_7(\mu)$ as
the scale $\mu$ is varied in the range $\mb/2 \leq \mu \leq 2 \mb$.
The partial next-to-leading-order, however, not unexpectedly 
shows a regularization scheme-dependence and its value is
also somewhat diminished compared to the leading-log estimates.
This reduces the branching ratio, and the (partial) 
next-to-leading-order estimates give
\brbgamaxs $= (1.9 \pm 0.2 \pm 0.5) \times 10^{-4}$ \cite{Apebkstar}.
The complete next-to-leading-order result is not yet available and hence
the estimates of the SM for \brbgamaxs ~are not completely
quantitative, although the inclusive rate (\ref{incl}) and the SM 
estimates \cite{ag5,Apebkstar} are in good agreement.
The sensitivity to new physics in \bgamaxs ~is therefore somewhat
entangled with the QCD corrections.
Despite this, the CLEO data on rare
$B$ decays have provided valuable constraints on the parameters of
models that are popular candidates for the
extensions of the SM, and in some cases
these constraints are very competitive compared to the
ones following from
direct and indirect searches \cite{CLEObsg}.

The determination of $\vert C_7(\mb)\vert$ from the inclusive branching
ratio \brbgamaxs ~is a prototype of the kind of analysis that we
would like to propose here to be carried out for the rare $B$ decays
in general and for the semileptonic decays \Bsell , in particular.
In the standard model, a measurement of
the radiative rare $B$ decay \bgamaxs ~mentioned above
and related ones, such as
\Bsell , \Bdell,
the corresponding exclusive decays $B \to (K, K^*, \pi,\rho,...)
\ell^+ \ell^-$, would lead to a determination of
$\mt$ and the CKM matrix element product $V_{ts}V_{tb}$ and
$V_{td}V_{tb}$. The purely leptonic decays $B_s^0 \to \ell^+ \ell^-$
and $B_d^0 \to \ell^+ \ell^-$
and the modes $B_s^0 \to \gamma \gamma$,
$B_d^0 \to \gamma \gamma$,
while having very different final
states and branching ratios --
and hence representing
experimentally very different propositions -- have more or less
the same information content in the SM as the decays mentioned
above (see, for
example, \cite{AGM93} and references cited therein).

The aim of this paper is to undertake first steps towards a
model-independent analysis of the FCNC electroweak
rare $B$ decays. Although the method of analysing
the data and the relevant theoretical framework
being presented here are developed for the inclusive decays
\bgamaxs ~and $B \to X_s + \ell^+ \ell^-$,
much of the general considerations being discussed apply also to
the corresponding exclusive decays such as \bksgam, ~\bksell ~and
$B \to K \ell^+ \ell^-$. Of course, the extraction of the 
short-distance physics in
terms of the Wilson coefficients of the dominant operators from the
data on exclusive decays would require the knowledge of the relevant
form factors. This may compromise the precision on the short-distance
part of the amplitudes in question at present, though advances in
computational methods for QCD may allow  quantitative
conclusions to be drawn from exclusive decays also.

Our analysis is based on the renormalization group (RG)-improved
treatment of
the effective Hamiltonian relevant for $B$ decays under consideration,
obtained by integrating out the top quark and the other heavy degrees
of freedom. The resulting Hamiltonian in the SM can be written as:
\beq
{\cal H}_{eff}(b \to sX) = -\frac{4G_F}{\sqrt{2}} \lambda_t
            \sum_{i=1}^{10} C_i(\mu) {\cal O}_i(\mu) ~.
\label{heff}
\eeq
Here $X$ stands for $q\bar{q}, ~\gamma$, gluon and $\ell^+ \ell^-$ and
$\lambda_t = V_{ts}^*V_{tb}$;
the operator basis contains dimension-5 and -6 operators and
is given in the appendix.

The problem of carrying out a model-independent analysis in FCNC
processes is evident from the SM expression for ${\cal H}_{eff}$
given above. In the most general case, which would also include
Left-Right-symmetric (LR) models, the operator basis consists of
20 operators, having 20 independent coefficients $C_i$. However,
as argued later, 
we shall limit the present analysis to left-handed fields only,
in which case there are still 10 operators to be considered. The
effective Hamiltonian then still involves 10 independent coefficients
$C_i$ and determining them experimentally is a monumental task.
It is obvious that some theoretical assumptions have to be used to
focus the attention on the more interesting
Wilson coefficients. In our opinion, the promising coefficients from
the stand-point of non-SM physics searches are
 the ones in which the electroweak loop effects
(penguins and boxes in the diagrammatic language) play the dominant role.

The analysis carried out here is restricted to the
models in which the effective Hamiltonian is of
the form given in (\ref{heff}). We show later how 
measurements can be analysed to determine the inadequacy of this
operator basis, should that happen.
Although by no means completely general,
the operator basis contained in ${\cal H}_{eff}$
encompasses most models of current theoretical
interest, which include, apart from the SM,
the 2HDM (both types I and II) and the MSSM. The effective Hamiltonian
in the four-generation models, as well as in models with anomalous trilinear
gauge and fermion couplings, can also be  written in the  above form.

The coefficients $C_i(M_W), i=1,\ldots,6$,
determine the non-leptonic $B$-decay rates and the $B$-hadron
lifetimes. Since data on $B$ decays
and the results obtained in the SM from (\ref{heff}), including
QCD effects, are in good
agreement with each other ($\pm 20\%$), there is not much room left 
for the first
six coeffficients involving the four-quark operators
to deviate from their SM values. So we fix $C_i(\Mw),i=1,\ldots,6$, to their
SM values. Beyond the leading-log approximation, these
receive corrections of order $\alpha_s (M_W)$
at the large scale $\mu = M_W$; they should be incorporated in
theoretical estimates consistently with other higher-order effects.
The coefficients of our interest are $C_7(\mu),...,C_{10}(\mu)$, since they
are generated at scale $\mu = M_W$ by electroweak loops (penguins and boxes).
They govern the physics of the rare $B$
decays $b \to s +g$, \bgamaxs , \Bsell ~and $B_s^0 \to \ell^+ \ell^-$.
We shall concentrate on them and show how to
extract these coefficients (both their signs and magnitudes) from data
on radiative and (semi)leptonic rare $B$ decays. These can be compared
with the SM and extensions of it to search for new physics.

The experimental quantities we consider in this paper are the
following:
\begin{itemize}
\item Inclusive radiative rare decay branching ratio \brbgamaxs ; 
\item Invariant dilepton mass distributions 
 $ d{\cal B}(\mbox{\Bsell})/d\hat{s}$; 
\item Forward-backward (FB) charge asymmetry
 ${\cal A}(\hat{s})$ in \Bsell .
\end{itemize}
The kinematic variables are defined as:
\begin{eqnarray}
u &=& (p_b-p_1)^2 - (p_b - p_2)^2, \nonumber\\
s &=& (p_1 + p_2)^2 , \nonumber \\
\hat{s} &=& \frac{s}{\mb^2} ,\nonumber\\
  w(s)&=&\sqrt{(s-(\mb + m_s )^2)
(s - (\mb -m_s )^2)}.
\label{kinvar}
\end{eqnarray}
where $p_b, ~p_1$ and $p_2$ denote, respectively, the momenta of the
$b$ quark ($=B$ hadron), $\ell^+$ and $\ell^-$.
The FB asymmetry
${\cal A}(\hat{s})$ is defined with respect to the angular variable
$z \equiv \cos \theta =u/w(s)$, where $\theta$ is the angle of the
$\ell^+$ with respect to the $b$-quark direction
in the centre-of-mass system of
the dilepton pair.
It is obtained by integrating the doubly differential
distribution $d^2 {\cal B} / (dz \, d\hat{s})$ \cite{AMM}:
\beq \label{asy}
{\cal A}(\hat{s}) \equiv \int\limits_0^1 dz \,
\frac{d^2{\cal B}}{dz \, d\hat{s}} (B \to X_s \ell^+ \ell^-) -
\int\limits_{-1}^0 dz \,
\frac{d^2{\cal B}}{dz \, d\hat{s}} (B \to X_s \ell^+ \ell^-).
\eeq

The rationale for concentrating on these measurements is the following.
We remark that the decay rate \brbgamaxs ~puts a bound
on the absolute value of the coefficient $C_7(\mu)$.
However, the radiative $B$ decay rate by itself
is not able to distinguish
between the solutions $C_7(\mu) > 0$ (holding in the SM) and the
solutions $C_7(\mu) < 0$, which, for example, are
also allowed in the MSSM 
as one scans over the allowed parameter space. We recall that
the invariant dilepton mass distribution and the forward-backward asymmetry
in \Bsell ~are sensitive to the sign and magnitude of $C_7(\mu)$
\cite{AMM,GSWll,Wylerbsll}.
This is easy to see in the approximation of neglecting
the $s$-quark mass, in which limit the differential branching fraction for
\bgamaxs ~can be written as:
\begin{eqnarray}
\frac{d{\cal B}( \mbox{\Bsell})}{d\hat{s}}
 &=& K (1- \hat{s})^2 \{ (\vert C_9^{eff}(\mu) \vert^2 +
  \vert C_{10}(\mu)
\vert^2)(1+ 2 \hat{s}) \nonumber\\
 & & + \vert C_7(\mu) \vert^2\frac{4}{\hat{s}}
(2 +  \hat{s}) +12 \mbox{ Re } (C_7(\mu) {C_9^{eff}}(\mu)) \} ,
\label{dbshat}
\end{eqnarray}
where $K$ is a constant and $C_9^{eff}(\mu)$ is defined as
\beq \label{c9eff}
C_9^{eff}(\mu)= C_9(\mu) + Y(\mu, \hat{s}) ~,
\eeq
where $Y(\mu, \hat{s})$ is a function involving the coefficients
$C_i(\mu); i=1,\ldots,6$, and it depends on the kinematic variable
$\hat{s}$ through the one-loop matrix element of the four-quark
operators. Long-distance effects would also be contained in
$Y(\mu, \hat{s})$; however, we shall not consider these here.

Consistent with our assumptions, we shall take the function $Y(\mu)$
from the SM and admit non-SM contributions only in  $C_7(\mu)$, 
$C_8(\mu)$. $C_9(\mu)$, and $C_{10}$.
{}From the above expression it is obvious that the
dilepton invariant-mass distribution is sensitive to the real part of
the product $C_7(\mu){C_9^{eff}}(\mu)$
in sign and magnitude. However, this distribution
by itself cannot determine the sign of $C_9^{eff}$ and $C_{10}$.
As the FB asymmetry and the dilepton rate are in general
independent quantities, they provide independent constraints on the
Wilson coefficients, and we argue that the FB asymmetry
${\cal A}(\hat{s})$
can be used to resolve these ambiguities.
In the limit $m_s =0$, the FB asymmetry has a simple form also:
\beq
{\cal A}(\hat{s})= -\frac{3}{2} K (1-\hat{s})^2  C_{10} (\mu)
\{C_9^{eff}(\mu) \hat{s} + 4 C_7(\mu)\} .
\label{asyms}
\eeq
The $\hat{s}$-dependence in ${\cal A}(\hat{s})$ now allows us to
disentangle the dependence on $C_7(\mu)$ and $C_9^{eff}(\mu)$.
Moreover, the FB asymmetry ${\cal A}(\hat{s})$
is directly proportional to the coefficient $C_{10}(\mu)$
and hence the asymmetry is very effective in constraining
$C_{10} (\mu)$.

The first
measurements will be done for partially integrated dilepton mass
distribution and
asymmetry. We discuss how to extract the Wilson coefficients from
these measurements.
A given integrated branching ratio ${\cal B} (\Delta s )$
over a given range in $ s $,
$\Delta  s $, and a
given value of $C_7(\mu)$
can  be analysed in terms of the contour plots drawn in the
Wilson-coefficients plane $C_9(\mu)$ and $C_{10}(\mu)$.
It is  obvious from the quadratic equation (\ref{dbshat}) that, for a given
$C_7(\mu)$, there is at most a
four-fold ambiguity in the signs of $C_9(\mu)$
and $C_{10}(\mu)$. In addition one may make use of the 
integrated FB asymmetry ${\cal A} (\Delta  s )$. For a fixed 
value of ${\cal A} (\Delta  s )$ one finds   
hyperbolae in the
$C_9$-$C_{10}$ plane, for given sign and magnitude of $C_7(\mu)$.
The intersection of the contours obtained from 
${\cal B}$(\Bsell ; $\Delta  s)$ and
${\cal A}(\Delta  s )$ then determines the signs and magnitudes
of all three Wilson coefficients
$C_7(\mu),C_9(\mu)$ and $C_{10}(\mu)$. We plot, in the 
$C_9(\mu)$-$C_{10}(\mu)$ plane, the contours that
are allowed from the solutions of the two algebraic equations for
${\cal B}(\Delta  s )$ and ${\cal A }(\Delta  s )$
for the two values of $C_7$, $C_7=\pm 0.3$, and two invariant-mass
intervals below the $J/\psi$ and above the $\psi^\prime$ mass
(see figs.~\ref{fig1}-\ref{fig4}), pointing out the
SM solution and the general solutions in non-SM cases. It is, however,
conceivable that there is no consistent solution of the two algebraic
equations for ${\cal B}(\Delta  s )$ and ${\cal A}(\Delta  s )$
in $B \to X_s \ell^+ \ell^-$ and the branching ratio $B \to X_s \gamma$
in terms of the coefficients $C_7(\mu),~C_9(\mu)$ and $C_{10}(\mu)$,
which would then indicate that the operator basis chosen in
describing the ${\cal H}_{eff}$ given above is incomplete.
The LR-symmetric models, which we have dropped from our
discussion, is a case in point.
Of course, given enough statistics the two distributions could be
used to determine the coefficients from the data fits   
and compared with the values in the
various models directly. We illustrate these
analysis techniques for some assumed values of the branching ratios and
asymmetry and representative values in the allowed
parameter space in various
models in terms of both the contour plots in the Wilson coefficient
space and the distributions themselves.

This paper is organized as follows. In section 2, we briefly review the
kinematics of the decays \bgamaxs ~and \Bsell , define the amplitudes
and the various distributions of interest.
In section 3, we perform a numerical study of the differential and
integrated branching ratio in \Bsell .
To illustrate the method, we have taken a representative
value for the effective Wilson coefficient, $C_7(\mb)=\pm 0.3$, which
lies within the presently allowed range for this quantity
from data on \bgamaxs .
We give the partially integrated branching ratio ${\cal B}(\Delta s)$
and the asymmetry ${\cal A}(\Delta   s)$, making the
dependence on the coefficients $C_9$ and $C_{10}$ explicit.
The resulting constraints on these coefficients from the ``low-dilepton
mass" and ``high-dilepton mass" regions are then displayed as contour
plots in the $C_9$-$C_{10}$ plane
for some illustrative values of the branching ratios and FB asymmetries.
The invariant
dilepton mass spectrum in \Bsell ~and  the FB asymmetry
are also shown here for some representative values of the Wilson
coefficients.
In section 4 we discuss the predictions of specific models, namely
the 2HDM (type I and type II) and the MSSM. These models allow a
calculation of the coefficients $C_7(\mu),C_9(\mu)$ and
$C_{10}(\mu)$ as a function of the model parameters. For the
case of the 2HDM we show  $C_7(\mu),C_9(\mu)$ and
$C_{10}(\mu)$ as a function of the charged Higgs mass $m_{H^+}$ for
representative values of the ratio of the two vacuum expectation
values $v_2/v_1$. As already noted \cite{GSWll}, the type-I 2HDMs
admit negative $C_7(\mu)$ solutions, as opposed to the positive $C_7(\mu)$
solution obtained in the SM. However, for the
type-I models the negative $C_7(\mu)$ solutions are 
excluded by present data, as will be shown in section 4.
To analyse the MSSM we vary the parameters of the model over the
experimentally allowed values, and the
resulting region in the
$C_{9}(\mu)$--$C_{10}(\mu)$ plane is shown.
In doing this, we have imposed the constraints on the
coefficient $C_7(\mu)$ as explained earlier. The restrictions on the
supersymmetric (SUSY) particle masses from present searches and the
anticipated reach of experiments at the Tevatron and LEP-II are
also imposed. The most interesting part of this exercise is that these
constraints do allow both the negative and positive
$C_7(\mu)$ solutions, which are consistent with the data on \bgamaxs ~but
admit values of the coefficients $C_9(\mu)$ and $C_{10}(\mu)$
sufficiently different from those of the SM, thereby yielding
very different differential distributions in the
decay \Bsell . These distributions are shown for some representative
values of the Wilson coefficients. Section 5 contains a summary.
To fix our notation we collect the relevant formulae for the
effective Hamiltonian, including the RG-improved
Wilson coefficients $C_i(\mu)$ in an appendix.

\section {Estimates of $\BBGAMAXS$,
 Invariant Dilepton Mass Distribution and Forward--Backward
 Asymmetry in \Bsell}

\indent
In the following we shall consider the inclusive decays
$B \to X_s \gamma$ and $B \to X_s \ell^+ \ell^-$, where $\ell$
is either electron or muon. It has been shown that inclusive
$B$ decays may be treated in a $1/m_b$ expansion, the leading term
of which is the free quark decay. The next-to-laeding effects are
of second order in $1/m_b$, i.e.~$O(m_s^2/\mb^2)$
\cite{Chayetal,Bigietal,Falketal}. This is true for total rates,
for partially integrated rates,
and also for decay distributions, as long as one is not too close
to the kinematic endpoint in which the energy release in the hadronic
subprocess becomes small.

We shall also not include QCD corrections to the free-quark decay
distributions,
aside from the leading logarithms, which are induced by the renormalization
group running. It is known that the measured inclusive lepton energy
distributions of the charged-current-induced semileptonic transitions
are well approximated by the partonic distributions
$b \to (c,u)  \ell^- \bar{\nu}_\ell$, including QCD corrections
\cite{Alipiet,ACCMM}. We expect  this for the flavour-changing
neutral current semileptonic transitions as well.

The second proviso is
that we shall concentrate on the short-distance contributions to the
decays \bgamaxs ~and ~\Bsell .
This again is not a drastic oversimplification, as it is known that the
long-distance contributions to the decay \bgamaxs ~are small
\cite{LMS,DPR,Desh} and those in the decays \Bsell ~are
dominantly present at and near the $J/\psi$ and $\psi'$ poles,
extending also to the region between them. These
contributions can be modelled using data on the
decays $B \to (J/\psi, \psi') X_s \to (\ell^+ \ell^-) X_s$, which
already exist and which are expected to become quite precise in the future.
We shall thus concentrate on the distributions in the regions away
from the resonances and consider the two regions
\begin{eqnarray}
\mbox{``Low dilepton-mass'':}~~~~~~~~~ 4\ml^2 &\leq& s \leq
 m_{J/\psi}^2 -\delta ~,\nonumber\\
\mbox{~~~``High dilepton-mass'':}~~~~~~~ m_{\psi'}^2+ \delta
 &\leq& s \leq
 s_{max},
\label{shatcuts}
\end{eqnarray}
where $\delta$ is a cut-off that can be matched with the experimental
cuts used in the analysis. In these two regions we expect only small
long-distance effects. Due to limitations in rates and
dilepton trigger requirements, we anticipate that the ``Low dilepton-mass
region" is accessible to $e^+e^-$ experiments (CLEO and $B$ factories)
and the ``High dilepton-mass region" typically to $B$ experiments
which will be carried out
with hadron beams (CDF, HERA-B, LHC). Given high enough luminosity,
this region can also be probed at the $B$ factories.

\subsection{ Decay rate for \bgamaxs }

The procedure for the computation of the Wilson coefficients in the SM
and extensions of it is by
now standard. They are obtained at a large scale $\mu^2=\Mw^2$
(in general at $\mu^2=m_Y^2$, where $m_Y$ stands
generically for the mass of the heavy degrees of freedom) by integrating
out degrees of freedom heavier than $\mu$ \cite{InamiLim}, and
are sensitive to the presence of new physics.
Then the renormalization group (RG) equations are used
to scale down these coefficients to the scale
that is typical for $B$-hadron decays, namely $\mu=O(\mb)$. In this
way large logarithms of the form $\alpha_s \ln M_W^2$ are shifted from the
matrix element into the Wilson coefficients.
The RG evolution of the Wilson coefficients $C_i(\mu)$
involves the $(10 \times 10)$ anomalous dimension
matrix, which has now been calculated at the one-loop level
\cite{Ciuchini,BLMM}. This matrix and the solutions of the RG equations
in terms of the effective coefficients $C_i(\mu)$
are given in the appendix.

Starting from the effective Hamiltonian as given in the appendix, one
finds that only one operator, namely ${\cal O}_7$ contributes at
tree level. The contribution of the QCD bremsstrahlung process \bsggam ~and
the virtual corrections to \bsgam ~have been calculated in $O(\alpha
\alpha_s)$ in ref.~\cite{ag1}. Expressed in terms of the inclusive
semileptonic branching ratio $BR(B \to X \ell \nu_\ell)$, one can
express the branching ratio \brbgamaxs ~as:
\begin{equation}
  \BBGAMAXS = 6 \frac{\alpha}{\pi}
 \frac{|\lambda_{t}|^2}{|V_{cb}|^2}
 \frac{|C_7(\mu)|^2 K(\mu)
                               }{f(m_c/m_b)[ 1-(2 \alpha_s) / (3 \pi)  
         h(m_c/m_b)]}\times {\cal B}_{sl} \ ,
    \label{e6}
\end{equation}
where
$$
f(r)=1-8r^2 +8r^6-r^8-24r^4\ln (r)
$$
is the phase-space function for $\Gamma(b \to c + \ell \nu_\ell)$
and ${\cal B}_{sl} = 10.5\%$ is the semileptonic branching fraction.
The function $h(r)$ accounts for QCD corrections to the semileptonic
decay and can be found, for example, in ref.~\cite{Alipiet}. It is a slowly
varying function of $r$ and, for a typical quark-mass ratio of
$r=0.35 \pm 0.05$, it has the value $h(r)=2.37 \mp 0.13$.
The contributions from the decays $b \to u  ~\ell ~\nu_\ell$
have been neglected in the denominator in (\ref{e6}) since they are
numerically  inessential because $\absvub\ll\absvcb$.

The inclusive decay width for \bgamaxs
~is dominantly contributed by the magnetic-moment
operator ${\cal O}_7$, hence
the rationale of factoring out its coefficient in the expression for
$\BBGAMAXS$ in Eq. (\ref{e6}).
Including $O(\alpha_s)$ corrections
brings to the fore other operators with their
specific Wilson coefficients. The
effect of these additional terms can be expressed in terms of
the function $K(\mu)$, which lumps together the effects of
bremsstrahlung corrections in the inclusive decay rate.
The function $K(\mu)$ has been computed in the SM in \cite{ag5}
taking into account the dominant corrections from $C_2$ and $C_8$ (the
coefficients of other operators are considerably smaller).
Typically, $0.79 \leq K(\mu) \leq 0.86$ for $\mb/2 \leq \mu \leq 2
\mb$. Since $C_8(\mu)$ receives possible non-SM contributions 
in the same way as 
$C_7(\mu)$, one should compute the $K$-factor and
incorporate these corrections into a full next-to-leading-order analysis.
However, we are not working beyond leading order
in this paper and so we ignore the $K$ factor.
Using the CLEO inclusive measurement 
\begin{equation} \label{bsgbounds}
\BBGAMAXS = (2.31 \pm 0.67)  \times 10^{-4} ,
\end{equation}
we get the following bound on $C_7(\mu)$:
\beq
  0.22 \leq \vert C_7(\mu) \vert \leq 0.30.
\eeq
Using, however, the 90\%-confidence-level range from the CLEO measurement
$\BBGAMAXS = (2.31 \pm 1.1)  \times 10^{-4}$ and the theoretical 
calculation for $\BBGAMAXS$ from \cite{ag5} we obtain
\begin{equation}
  0.19 \leq \vert C_7(\mu) \vert \leq 0.32.
\end{equation}
This range, in our opinion, adequately reflects the present uncertainties. 
When not stated otherwise, we shall fix $\vert C_7(\mu) \vert= 0.3 $, 
which is in comfortable agreement with the CLEO data.
\subsection{ Decay distributions in \Bsell }
Using ${\cal H}_{eff}$ given in
(\ref{heff}), one obtains for the dilepton invariant mass
distribution 
\begin{eqnarray} \label{AAA}
{d{\cal B} \over d\s}
&=& {\cal B}_{sl} \frac{\alpha^2}{4 \pi^2} \frac{\lambda_t}{\absvcb}^2
 \frac{1}{f(m_c/m_b)}
              \hat{w} (\hat{s}) \left[ \vphantom{\frac{1}{1}}
\left( |C_9 + Y(\hat{s})|^2
 + C_{10}^2 \right) \alpha_1 (\hat{s},\hat{m}_s)
\right. \\ \nonumber
&& \left. + \frac{4}{\hat{s}} C_7^2 \alpha_2 (\hat{s},\hat{m}_s)
+ 12 \alpha_3 (\hat{s},\hat{m}_s) C_7 (C_9 + \mbox{ Re }Y(s)) \right] ,
\label{dbrs}
\end{eqnarray}
where the auxiliary functions are defined as follows:

\begin{eqnarray}
\alpha_1 (\hat{s},\hat{m}_s) &=& - 2 \hat{s}^2 + \hat{s} (1+ \hat{m}_s^2)
                           +(1-\hat{m}_s^2)^2     \\
\alpha_2 (\hat{s},\hat{m}_s) &=& -(1+ \hat{m}_s^2) \hat{s}^2
                           - (1+14 \hat{m}_s^2+\hat{m}_s^4) \hat{s}
                           +2 (1+ \hat{m}_s^2)(1-\hat{m}_s^2)^2  \\
\alpha_3 (\hat{s},\hat{m}_s) &=& (1-\hat{m}_s^2)^2 - (1+ \hat{m}_s^2)
                           \hat{s} \\
Y(\hat{s}) &=& g(m_c/m_b,\hat{s})
(3C_1 + C_2 + 3C_3 + C_4 + 3 C_5 + C_6) \\
 && - \frac{1}{2} g(1,\hat{s}) ( 4C_3 + 4C_4 + 3C_5 + C_6) \nonumber \\
     && - \frac{1}{2} g(0,\hat{s}) (C_3 + 3C_4) + \Delta C_9 \nonumber
\end{eqnarray}
and $g(z,\hat{s})$ is the one-loop function given in the appendix.
Furthermore,
the constant $\Delta C_9$ depends on the scheme, which is chosen in the
evaluation of the one-loop matrix elements of the operators
${\cal O}_1 \cdots {\cal O}_6$. It contains also a large logarithm
$\ln (M_W^2/ m_b^2)$, which is not due to QCD effets but rather comes
from the one-loop matrix elements of ${\cal O}_1 \cdots {\cal O}_6$.
In the $\overline{\rm MS}$ scheme one obtains \cite{GSWll}
\begin{equation}
\Delta C_9 = \frac{4 \pi}{\alpha_s (M_W)} \left\{ \frac{4}{33}
\left[
1-\left(\frac{\alpha_s(M_W)}{\alpha_s(m_b)}\right)^{-11/23}\right]
-\frac{8}{87}
\left[
1-\left(\frac{\alpha_s(M_W)}{\alpha_s(m_b)}\right)^{-29/23}\right]
\right\} ,
\end{equation}
which generates the correct logarithms in the limit $\alpha_s \to 0$.

The corresponding differential asymmetry as defined in (\ref{asy}) is
\begin{eqnarray}
{\cal A} (\hat{s}) &=& - {\cal B}_{sl} \frac{3 \alpha^2}{8 \pi^2}
                      \frac{1}{f(m_c/m_b)} \hat{w}^2 (\hat{s})
C_{10} \left[ \hat{s} ( C_9 + \mbox{ Re } Y(\hat{s})) +
4 C_7 (1 + \hat{m}_s^2) \right] .
\label{dasym}
\end{eqnarray}

In the subsequent section, we shall be evaluating the partial branching
ratio ${\cal B}(\Delta  s ) $ and partial FB asymmetry
${\cal A} (\Delta  s ) $,
where $\Delta  s $ defines an interval in the dilepton invariant
mass. All numerical
calculations are done with a non-zero value for the
$s$-quark mass, $m_s = 500$~MeV.
The asymmetry in the dilepton angular distribution in the SM
can be qualitatively understood as follows.
The decays \Bsell ~occur through $\gamma$ , $Z$ and
${W^+ W^-}$
exchange diagrams.
For small $\mt$
$(\mt / \Mw \ll 1)$ the photon
contribution dominates and the vector-like interactions to the
leptonic current remain substantial; consequently, the asymmetry is
small. However, for $\mt / \Mw \geq 2$, as suggested by the CDF
value for $\mt$,
the contribution from the $Z$-exchange diagrams becomes important and the
coefficient of the left-handed leptonic current grows as $\mt^2$,
leading to a large asymmetry.

\section{Analysis of the Decays $B \to X_s \gamma$ and
         $B \to X_s \ell^+ \ell^-$ }
In this section we shall discuss how the Wilson
coefficients appearing in the effective Hamiltonian may be extracted
from the experimental information. We shall assume that all the matrix
elements are normalized at the scale $\mu \sim m_b$, the mass of the
$b$ quark and hence the decay distributions are given in terms of 
the Wilson coefficients at the scale $m_b$.
The SM makes specific predictions for these
coefficients (modulo perturbative QCD uncertainties), but if
there is physics beyond the SM, these
coefficients will in general be modified.

We will somewhat elaborate on this point.
A specific model provides the set of Wilson coefficients
at high scales, which we shall choose to be the scale of the weak
bosons $\mu =M_W$. Furthermore, we shall integrate out
heavy degrees of freedom at the same scale $\mu = M_W$; this
procedure introduces an uncertainty due to the difference in the
masses of the heavy degrees of freedom, as for example arising from
$m_t \simeq 2 M_W$. However, since the QCD coupling constant is small
at these very high scales and does not
appreciably change between these thresholds, it is a reasonably accurate
approximation to neglect QCD corrections for scales above
$\mu = M_W$. Starting from this scale, the Wilson coefficients
are obtained from the solution of the renormalization group equations
at the scale $\mu \sim m_b$, where we use the one-loop result for
the anomalous dimensions and the beta function (see appendix).

In order to determine the sign of $C_7$ and the other two coefficients
$C_9$ and $C_{10}$, one has to study the decay distributions and rates
in $B \to X_s \ell^+ \ell^-$, where $\ell$ is either electron or muon.
As already discussed,
these decays are sensitive to the sign of $C_7$, and to $C_9$ and $C_{10}$.
The first experimental information available in the decay
$B \to X_s \ell^+ \ell^-$ will be a measurement of the branching
fraction in a certain kinematic region of the invariant mass $s$ of the
lepton pair. In order to minimize long-distance effects we shall
consider the kinematic regime for $s$ below the $J/\psi$ mass
(low invariant mass) and for $s$ above the mass of the $\psi '$
(high invariant mass). Integrating (20) over these regions for
the invariant mass one finds\footnote{
    In performing the integrations over $\Delta s$ we have set
    the resolution parameter $\delta$ to zero, since we do not consider any
    long-distance contribution. The long-distance contribution peaks
    strongly at the $J/\psi$ and $\psi '$ and $\delta$ has to be several
    times the width of these resonances in order to avoid large 
    long-distance effects. However, calculating only the short-distance
    part one may safely neglect $\delta$, since the short-distance
    contribution is flat in this region.}
\begin{equation} \label{branch}
{\cal B} (\Delta s) = A(\Delta s) \left( C_9^2 + C_{10}^2 \right)
                      + B(\Delta s) C_9 + C(\Delta s) ,
\label{bds}
\end{equation}
where $A$, $B$ and $C$ are fixed in terms of the Wilson coefficients
$C_1 \cdots C_6$ and $C_7$. We derive from (\ref{AAA}):
\begin{eqnarray}
A(\Delta s) &=& {\cal B}_{sl} \frac{\alpha^2}{4 \pi^2}
\frac{1}{f(m_c/m_b)}
                \int\limits_{\Delta s} d\hat{s} \, \hat{u} (\hat{s})
        \alpha_1 (\hat{s},\hat{m}_s) \\
B(\Delta s) &=& {\cal B}_{sl} \frac{\alpha^2}{4 \pi^2}
\frac{1}{f(m_c/m_b)}
\int\limits_{\Delta s} d\hat{s} \, \hat{u} (\hat{s}) \left[
2 \alpha_1 (\hat{s},\hat{m}_s) \mbox{ Re } Y(\hat{s}) + 12 C_7
            \alpha_3 (\hat{s},\hat{m}_s)\right]
\\ \label{C}
C(\Delta s) &=& {\cal B}_{sl} \frac{\alpha^2}{4 \pi^2}
\frac{1}{f(m_c/m_b)}
\int\limits_{\Delta s} d\hat{s} \, \hat{u} (\hat{s}) \left[
\alpha_1 (s,\hat{m}_s)
\left\{ (\mbox{ Re } Y(\hat{s}) )^2 + (\mbox{ Im } Y(\hat{s}) )^2
\right\} \nonumber
\vphantom{\frac{4}{s}}   \right. \\
&& \left. + \frac{4}{s} |C_7|^2 \alpha_2 (s,\hat{m}_s)
+ 12 \alpha_3 (s,\hat{m}_s) C_7 \mbox{ Re } Y(\hat{s}) \right] ,
\end{eqnarray}
where the auxiliary functions $\alpha_i$, $i=1,2,3$, are as given
above.

In our analysis we keep the values for $C_1 \cdots C_6$
and the modulus of $C_7$ fixed and hence $A(\Delta s)$, $B(\Delta s)$
and
$C(\Delta s)$ may be calculated for the two invariant-mass ranges of
interest. For the numerical analysis we use $m_b = 4.7$ GeV,
$m_c = 1.5$ GeV, $m_s = 0.5$ GeV. The
values
for the Wilson coefficients $C_1, \ldots , C_6$ are given in the appendix.
The resulting coefficients $A$, $B$, and $C$ are listed in Table
\ref{tab1}.

\begin{table}
\begin{center}
\begin{tabular}{| c | c | c | c | c | c |}
\hline
 $\Delta s$ & $C_7$   & $A(\Delta s) / 10^{-8}$
                              & $B(\Delta s) / 10^{-8}$
                              & $C(\Delta s) / 10^{-8}$
                              & $C(\Delta s) / 10^{-8}$  \\
 & & & & $\ell = e$ & $\ell = \mu$ \\
\hline \hline
$4m_\ell^2 < s < m_{J/\psi}^2$& $+0.3$ & 2.86 & $-$5.76 & 84.1 & 76.6 \\
$4m_\ell^2 < s < m_{J/\psi}^2$& $-0.3$ & 2.86 & $-$20.8 & 124  &  116 \\
$m_{\psi'}^2 < s < (1-m_s^2)$& $+0.3$ & 0.224 & $-$0.715 & 0.654 & 0.654 \\
$m_{\psi'}^2 < s < (1-m_s^2)$& $-0.3$ & 0.224 & $-$1.34  & 2.32  & 2.32 \\
\hline
\end{tabular}
\end{center}
\caption{Values for the coefficients $A(\Delta s)$, $B(\Delta s)$
         and $C(\Delta s)$ for the decay $B \to X_s \ell^+ \ell^-$.
         }
\label{tab1}
\end{table}

Inspection of (\ref{C}) shows that the integrand for $C(\Delta s)$
behaves as $1/s$ for small values of $s$, leading to a logarithmic
dependence of $C(\Delta s)$ on the lepton mass
for the case of the low invariant mass
region. In fact, this is the only point
where the masses of the leptons enter our analysis, and from this one
may obtain the corresponding coefficients for $\ell = \mu$:
\begin{equation} \label{lowercut}
C(4m_\mu^2 < s < m_{J/\psi}^2) =  C(4m_e^2 < s < m_{J/\psi}^2)
- 8 |C_7|^2 (1+ \hat{m}_s^2)(1-\hat{m}_s^2)^3
  \ln \left(\frac{m_\mu^2}{m_e^2}\right) .
\end{equation}
Of course one may apply (\ref{lowercut}) also to obtain $C$ for
any other lower cut $s_0$ on the lepton invariant mass, as long as
$s_0 \ll m_{J/\psi}^2$.

For a measured branching fraction ${\cal B} (\Delta s)$, one can solve
the above equation for ${\cal B }(\Delta s)$, obtaining
concentric
circles in the $C_9$-$C_{10}$ plane, with their centre lying at
$C_9^* =  B(\Delta s)/(2 A(\Delta s))$ and $ C_{10}^* = 0$.
The radius $R$ of these circles is proportional to

\begin{equation}
R = \sqrt{{\cal B}(\Delta s)-{\cal B}_{min}(\Delta s)} ,
\end{equation}
where the minimum branching fraction

\begin{equation}
{\cal B}_{min}(\Delta s) = C(\Delta s) -
            \frac{B^2(\Delta s)}{4 A(\Delta s)}
\end{equation}
is determined mainly by the present data on $B \to X_s \gamma$,
i.e.~by $|C_7|$. For the cases of interest one obtains, with the help
of Table \ref{tab1}:
\begin{eqnarray}
{\cal B}_{min} (4m_e^2 < s < m_{J/\psi}^2) &=& \left\{
\begin{array}{l} 8.1 \times 10^{-7} \mbox{ for } C_7 = 0.3 \\
                 8.6 \times 10^{-7} \mbox{ for } C_7 = -0.3 , 
\end{array} \right.\\ 
{\cal B}_{min} (m_{\psi'}^2 < s < (1-m_s^2) ) &=& \left\{
\begin{array}{l} 8.5 \times 10^{-10} \mbox{ for } C_7 = 0.3 \\
                 3.0 \times 10^{-9} \mbox{ for } C_7 = -0.3 . \end{array}
\right.
\end{eqnarray}

Note that ${\cal B} (\Delta s)$
is an even function of $C_{10}$, so one is not able to fix the
sign of $C_{10}$ from a measurement of ${\cal B} (\Delta s)$ alone.

To further pin down the Wilson coefficients,
one could perform a measurement of the forward-backward
asymmetry ${\cal A}$, which has been defined above. The asymmetry is
an odd function of $C_{10}$, and for a fixed value of the total
branching ratio in this kinematic region one obtains, from integrating
over a range $(\Delta s)$:
\begin{equation} \label{FB}
{\cal A} (\Delta s) = C_{10}
\left(\alpha (\Delta  s) C_9 + \beta(\Delta  s) \right) ,
\end{equation}
where

\begin{eqnarray}
\alpha (\Delta  s) &=& - {\cal B}_{sl} \frac{3 \alpha^2}{8 \pi^2}
                      \frac{1}{f(m_c/m_b)}
\int\limits_{\Delta  s} d\hat{s} \, \hat{u}^2 (\hat{s}) \hat{s} \\
\beta (\Delta s) &=& - {\cal B}_{sl} \frac{3 \alpha^2}{8 \pi^2}
                      \frac{1}{f(m_c/m_b)}
\int\limits_{  \hat s} d\hat{s} \, \hat{u}^2 (\hat{s})
\left[ \hat{s} \mbox{ Re } Y(\hat{s}) + 4 C_7 (1 + \hat{m}_s^2) \right] .
\end{eqnarray}

For a fixed value of ${\cal A} (\Delta s)$, one obtains
hyperbolic curves in the $C_9 $-$C_{10}$ plane; like
the coefficients $A$, $B$ and $C$, the parameters $\alpha$ and $\beta$ are
given in terms of the Wilson coefficients $C_1 \cdots C_6$ and $C_7$,
and
the kinematic region of $s$ considered; their values are presented in
Table \ref{tab2}.

\begin{table}
\begin{center}
\begin{tabular}{| c | c | c | c | }
\hline
 $\Delta s$ & $C_7$ & $\alpha(\Delta s)/10^{-9}$
                            & $\beta (\Delta s)/10^{-9}$ \\
\hline \hline
$4m_\ell^2 < s < m_{J/\psi}^2$& $+0.3$ & $-$6.08 & $-$24.0  \\
$4m_\ell^2 < s < m_{J/\psi}^2$& $-0.3$ & $-$6.08 &   55.4   \\
$m_{\psi'}^2 < s < (1-m_s^2)$& $+0.3$ & $-$0.391 & 0.276 \\
$m_{\psi'}^2 < s < (1-m_s^2)$& $-0.3$ & $-$0.391 & 1.37  \\
\hline
\end{tabular}
\end{center}
\caption{Values for the coefficients $\alpha(\Delta s)$and
         $\beta(\Delta s)$.}
\label{tab2}
\end{table}

\begin{figure}[p]

   \vspace{-0.5cm}
   \epsfysize=11cm
   \centerline{\epsffile{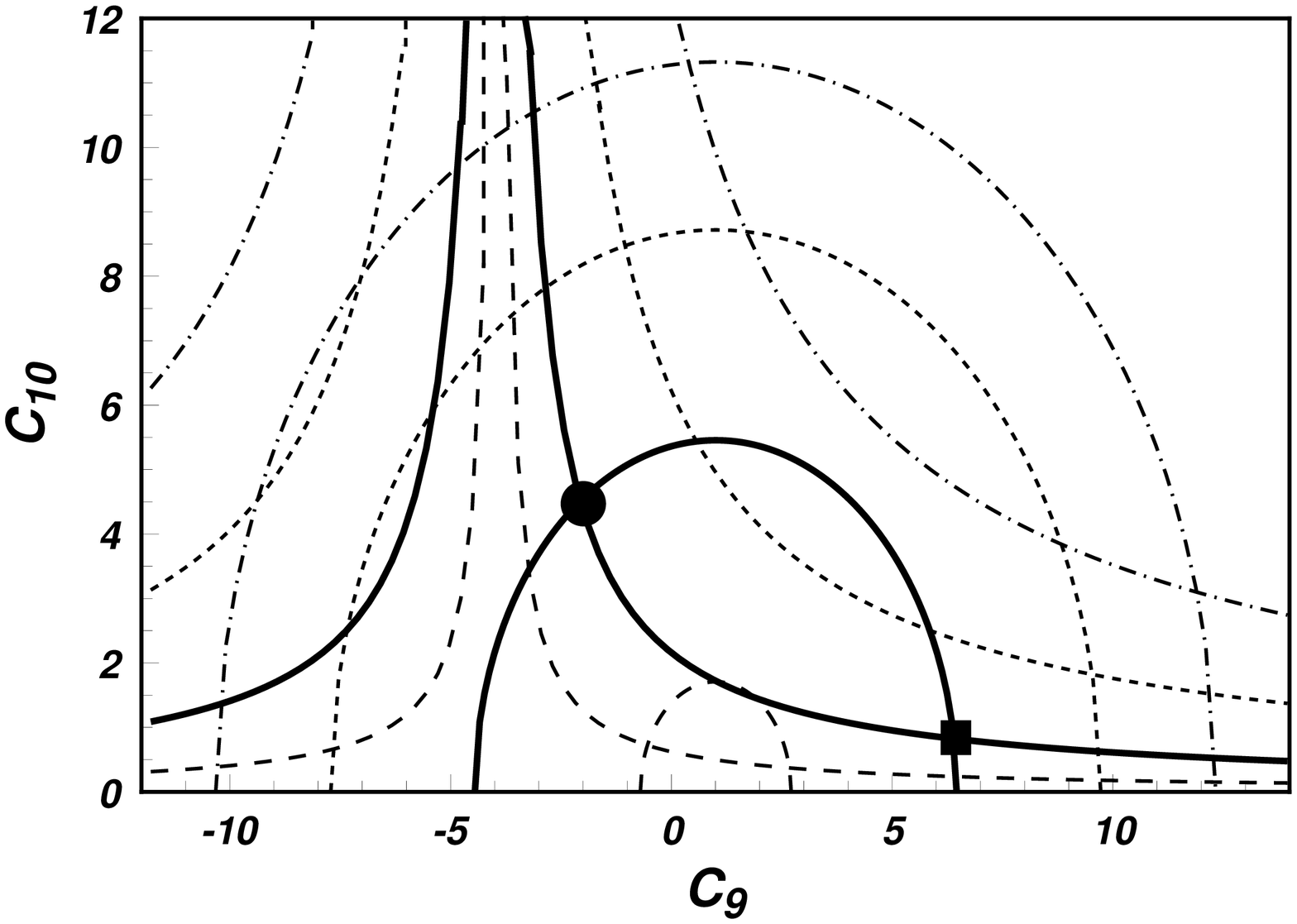}}
   \centerline{\parbox{16cm}{\caption{\label{fig1}
Contour plots of
${\cal B} (\Delta s) $ and ${\cal A} (\Delta s) $ in the $C_9$-$C_{10}$ 
plane
for the low-invariant-mass region $4 m_\ell^2 < s < m_{J/\psi}^2$ and
$C_7 = 0.3$. The circles correspond to fixed values of ${\cal B}$:
${\cal B} = 5.6 \times 10^{-6}$ (solid curve),
${\cal B} = 3.0 \times 10^{-6}$ (long-dashed curve),
${\cal B} = 1.0 \times 10^{-5}$ (short-dashed curve),
${\cal B} = 1.5 \times 10^{-5}$ (dash-dotted curve).
The left branches of the hyperbolae correspond to positive
values of ${\cal A}$:
${\cal A} = 1.7 \times 10^{-7}$ (solid curve),
${\cal A} = 5.0 \times 10^{-8}$ (long-dashed curve),
${\cal A} = 5.0 \times 10^{-7}$ (short-dashed curve),
${\cal A} = 1.0 \times 10^{-6}$ (dash-dotted curve).
The right branches of the hyperbolae correspond to negative
values of ${\cal A}$:
${\cal A} = -1.41 \cdot 10^{-8}$ (solid curve),
${\cal A} = -5.0 \cdot 10^{-9}$ (long-dashed curve),
${\cal A} = -3.0 \cdot 10^{-8}$ (short-dashed curve),
${\cal A} = -6.0 \cdot 10^{-8}$ (dash-dotted curve).
For negative values of $C_{10}$, the figure is simply reflected
with ${\cal A} \to - {\cal A}$.
The solid dot
indicates the SM values for $C_9$ and $C_{10}$.
The solid square is another allowed solution resulting from
the SM values of
${\cal B}$ and ${\cal A}$.}}}
\end{figure}

\begin{figure} 

   \vspace{-0.5cm}
   \epsfysize=11cm
   \centerline{\epsffile{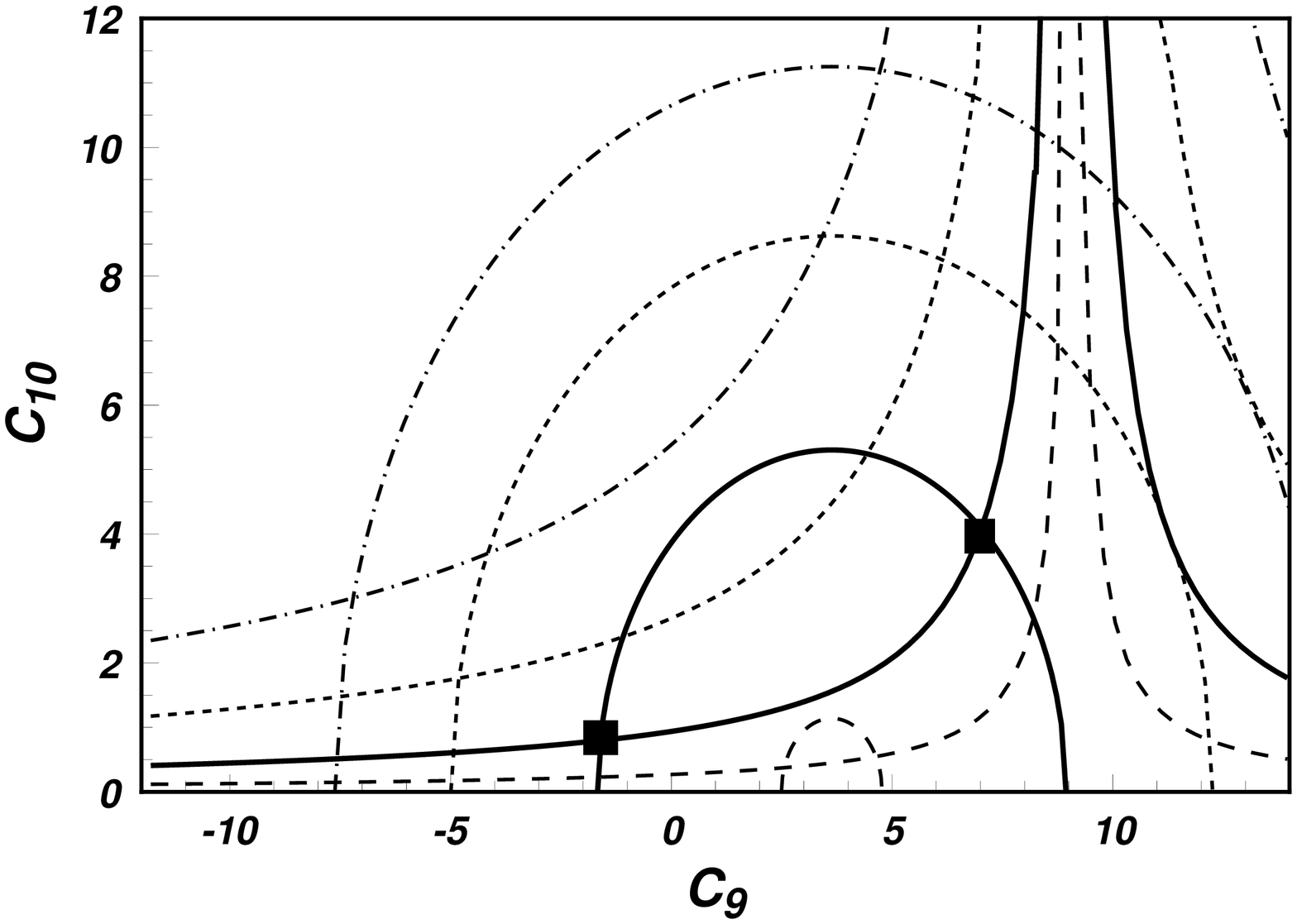}}
   \centerline{\parbox{16cm}{\caption{\label{fig2}
Same as in fig. 1, but for $C_7 = -0.3$. 
}}}
\end{figure}

\begin{figure}[p]
   \vspace{-0.5cm}
   \epsfysize=11cm
   \centerline{\epsffile{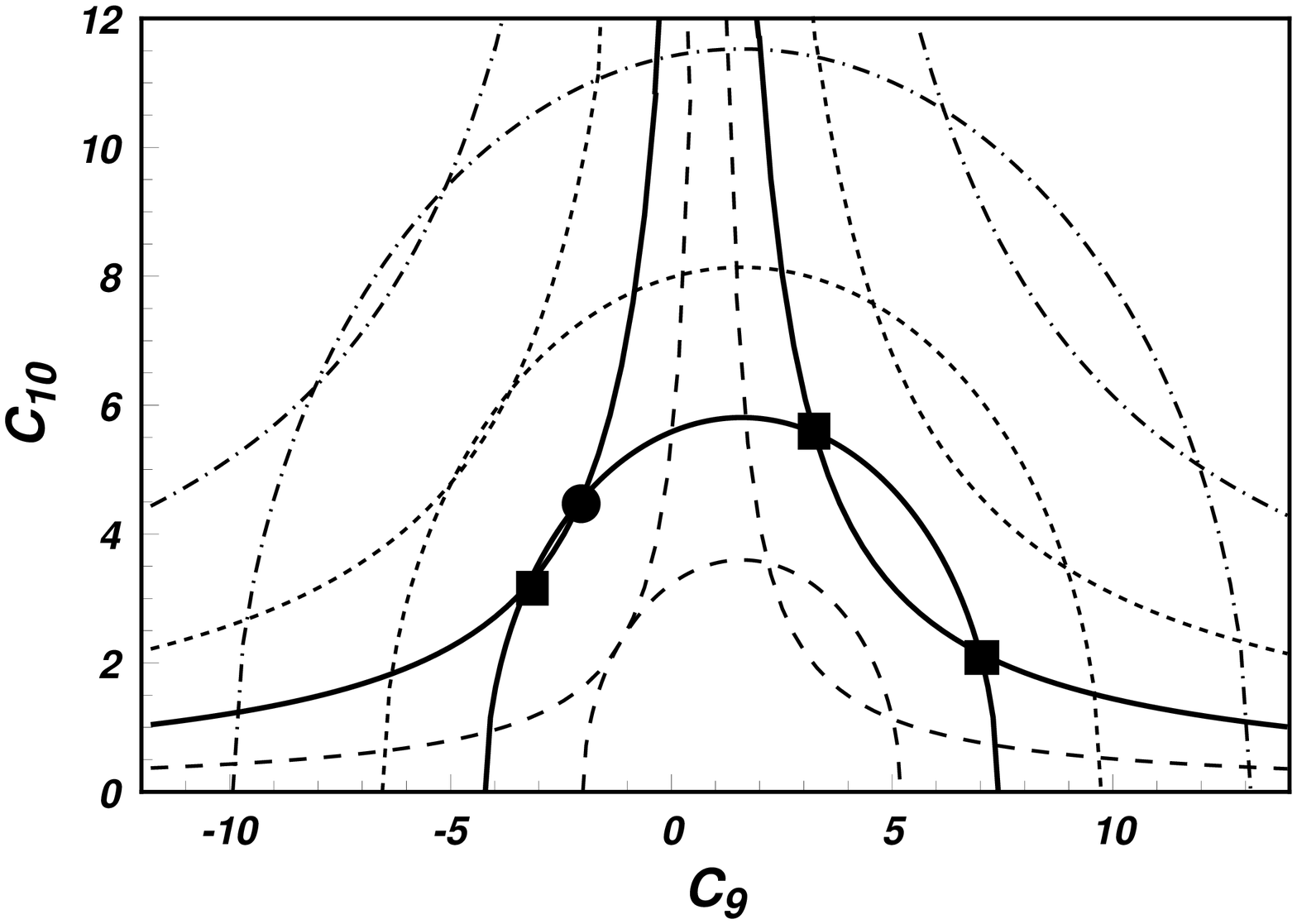}}
   \centerline{\parbox{16cm}{\caption{\label{fig3}
Contour plots of
${\cal B} (\Delta s) $ and ${\cal A} (\Delta s) $ in the $C_9$-$C_{10}$ 
plane
for the high-invariant-mass region $m_{\psi '}^2 < s < (1-m_s)^2$ and
for $C_7 = 0.3$.
The circles correspond to fixed values of ${\cal B}$:
${\cal B} = 2.56 \times 10^{-7}$ (solid curve),
${\cal B} = 1.0 \times 10^{-7}$ (long-dashed curve),
${\cal B} = 5.0 \times 10^{-7}$ (short-dashed curve),
${\cal B} = 1.0 \times 10^{-6}$ (dash-dotted curve).
The left branches of the hyperbolae correspond to positive
values of ${\cal A}$:
${\cal A} = 1.41 \times 10^{-8}$ (solid curve),
${\cal A} = 5.0 \times 10^{-9}$ (long-dashed curve),
${\cal A} =  3.0 \times 10^{-8}$ (short-dashed curve),
${\cal A} =  6.0 \times 10^{-8}$ (dash-dotted curve).
The right branches of the hyperbolae correspond to negative
values of ${\cal A}$:
${\cal A} = -1.41 \times 10^{-8}$ (solid curve),
${\cal A} = -5.0 \times 10^{-9}$ (long-dashed curve),
${\cal A} = -3.0 \times 10^{-8}$ (short-dashed curve),
${\cal A} = -6.0 \times 10^{-8}$ (dash-dotted curve).
For negative values of $C_{10}$, the figure is simply reflected
with ${\cal A} \to - {\cal A}$.
The solid dot
indicates the SM Values for $C_9$ and $C_{10}$.
The solid squares are other allowed solutions resulting from
the SM values of
${\cal B}$ and ${\cal A}$.}}}

\end{figure}

\begin{figure} 
   \vspace{-0.5cm}
   \epsfysize=11cm
   \centerline{\epsffile{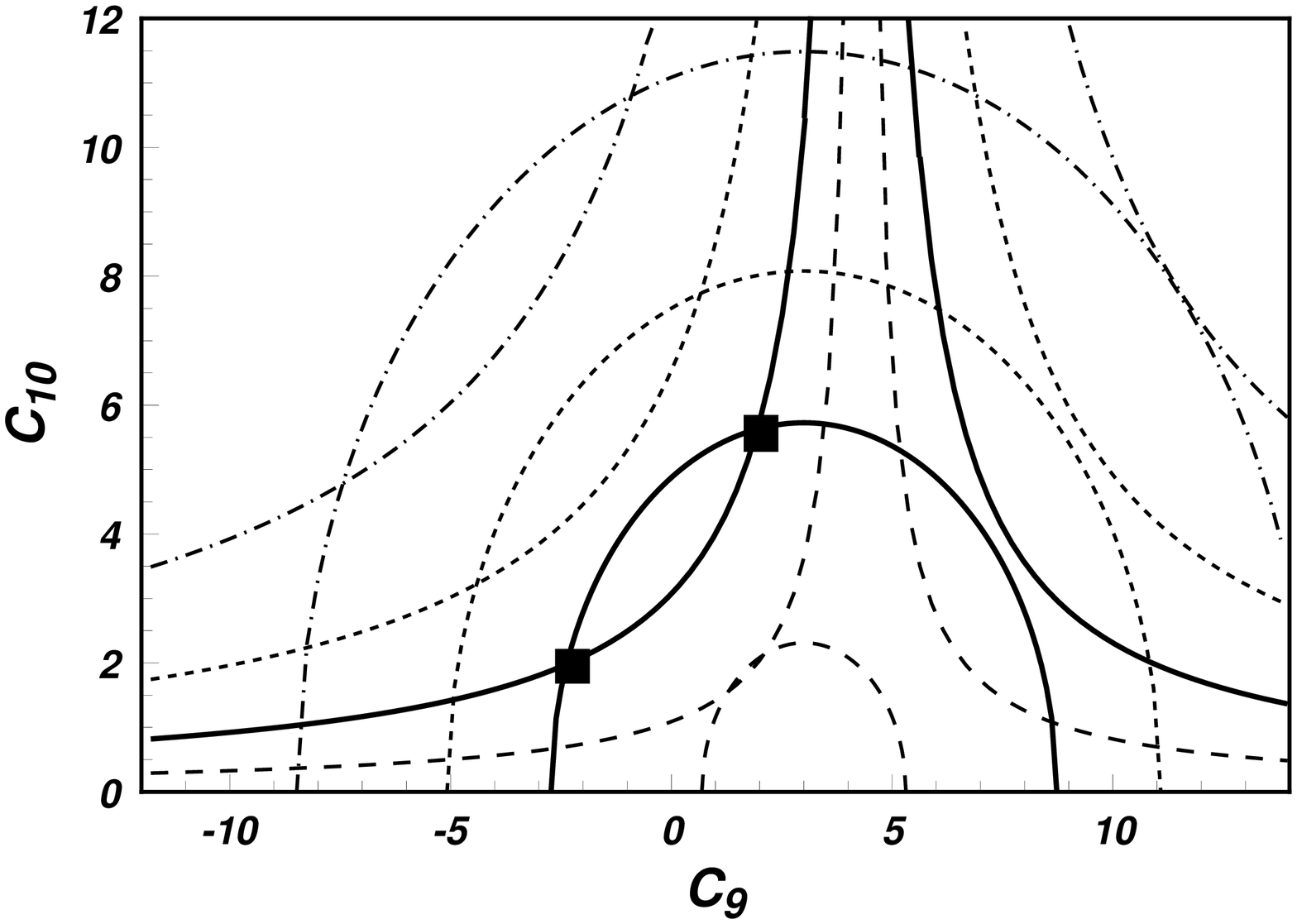}}
   \centerline{\parbox{16cm}{\caption{\label{fig4}
Same as in fig. 3, but
for $C_7 = - 0.3$.}}}
\end{figure}

Given the two experimental inputs, the branching fraction
${\cal B}(\Delta s)$ and the corresponding asymmetry ${\cal A}(\Delta s)$,
one obtains a fourth-order equation for the Wilson coefficients
$C_9$ and $C_{10}$, which admits in general four solutions.
In figs. \ref{fig1}-\ref{fig4} we plot the contours for a fixed
value for the branching fraction ${\cal B} (\Delta s)$ and the
FB asymmetry
${\cal A} (\Delta  s)$. Since ${\cal B} (\Delta s)$
is an even function of
$C_{10}$ and ${\cal A} (\Delta s)$ is
an odd one, we only plot positive values for $C_{10}$. 
The asymmetry vanishes for $C_{10} = 0 $, but also for 
$C_9 = - \beta(\Delta s) / \alpha(\Delta s)$. The two lines 
$C_{10} = 0 $ and $C_9 = - \beta(\Delta s) / \alpha(\Delta s)$ 
divide the $C_9$-$C_{10}$ plane into four quadrants, in which the 
asymmetry has a definite sign. Reflecting the hyperbolae on the line    
$C_{10} = 0 $ or $C_9 = - \beta(\Delta s) / \alpha(\Delta s)$ results 
in a sign change of the asymmetry.

Figures~\ref{fig1} and \ref{fig2} show
the contours in the $C_9(\mu) $-$C_{10}(\mu)$
plane for the low-invariant-mass region $4m_e^2 < s < m_{J/\psi}^2$, 
and figs.~\ref{fig3}
and \ref{fig4} are for the high-invariant-mass region
$m_{\psi'}^2 < s < (1-\hat{m}_s)^2$.
Figures~\ref{fig1} and \ref{fig3}
are obtained for  $C_7(\mu) = 0.3$, while
figs.~\ref{fig2} and \ref{fig4} are for $C_7(\mu) = -0.3$.
The possible solutions
for $C_9$ and $C_{10}$ are given by the intersections of the circle
corresponding to the measured branching fraction and the hyperbola,
corresponding to the measured asymmetry. Assuming
the SM values for both ${\cal B}(\Delta s)$ and ${\cal A}(\Delta s)$,
one obtains the solid lines in figs. \ref{fig1}-\ref{fig4}. The
possible
solutions in this case are represented by solid dots (SM solutions)
and solid squares (other non-SM possible solutions).

{}From the figures one reads off that for the SM values of ${\cal B}$ and
${\cal A}$  one has
more than one solution for the coefficients $C_9$ and $C_{10}$,
but the ambiguity may in general be resolved by measuring both
the low and the high invariant mass regions.

However, there is in principle also the possibility that the
equations do not have a solution for $C_9$ and $C_{10}$. This is the
case, for example,
when the asymmetry is large and the branching fraction small, in
which case the hyperbola may not intersect with the corresponding
circle any more. If this happens one has to conclude that the present
analysis is not complete; in other words, the operator basis we started
from is not complete and physics beyond the SM will be
present
in the form of additional operators such as right-handed currents.

\begin{figure}[p]

   \vspace{-0.5cm}
   \epsfysize=9cm
   \centerline{\epsffile{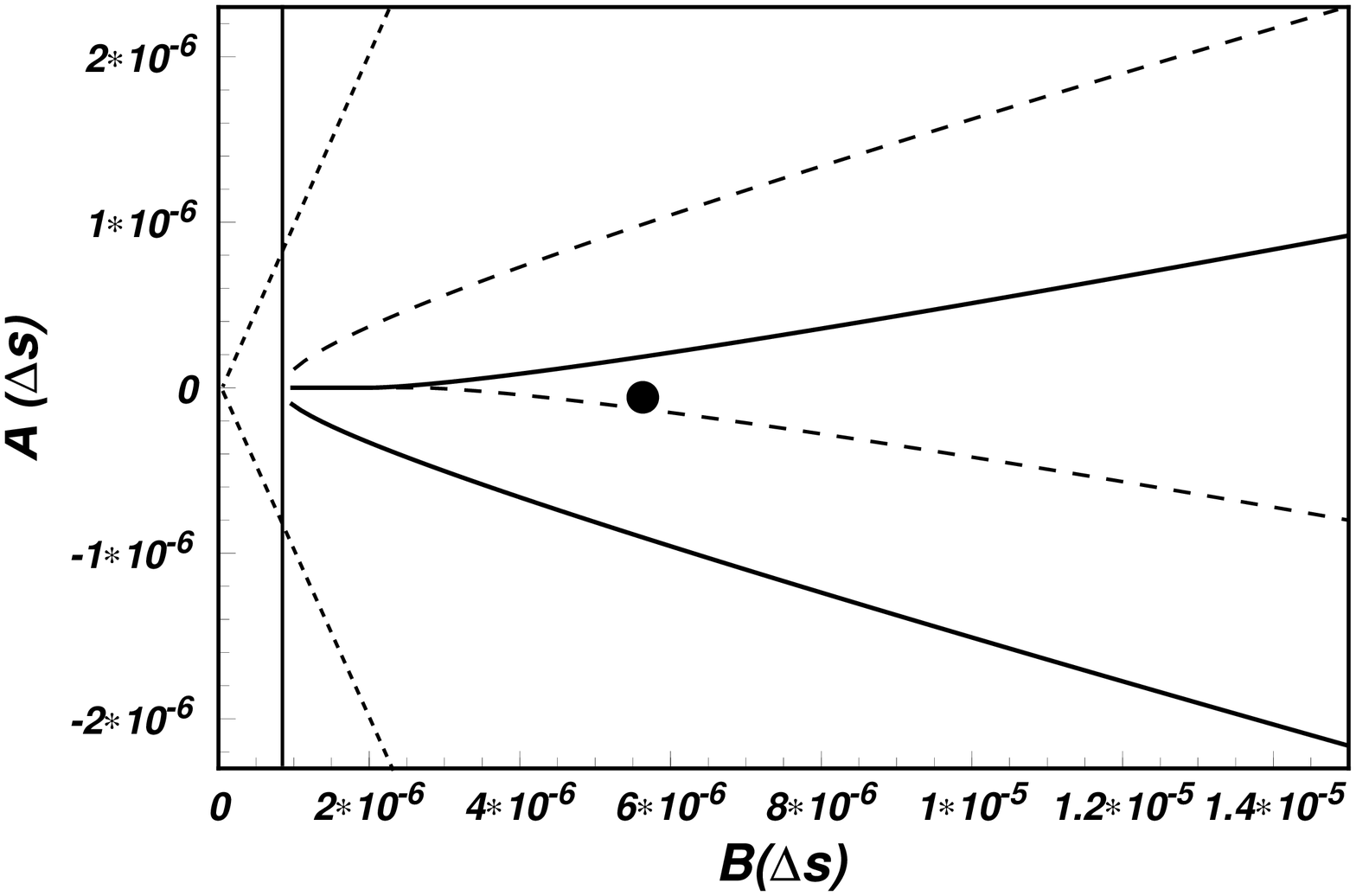}}
   \epsfysize=9cm
   \centerline{\epsffile{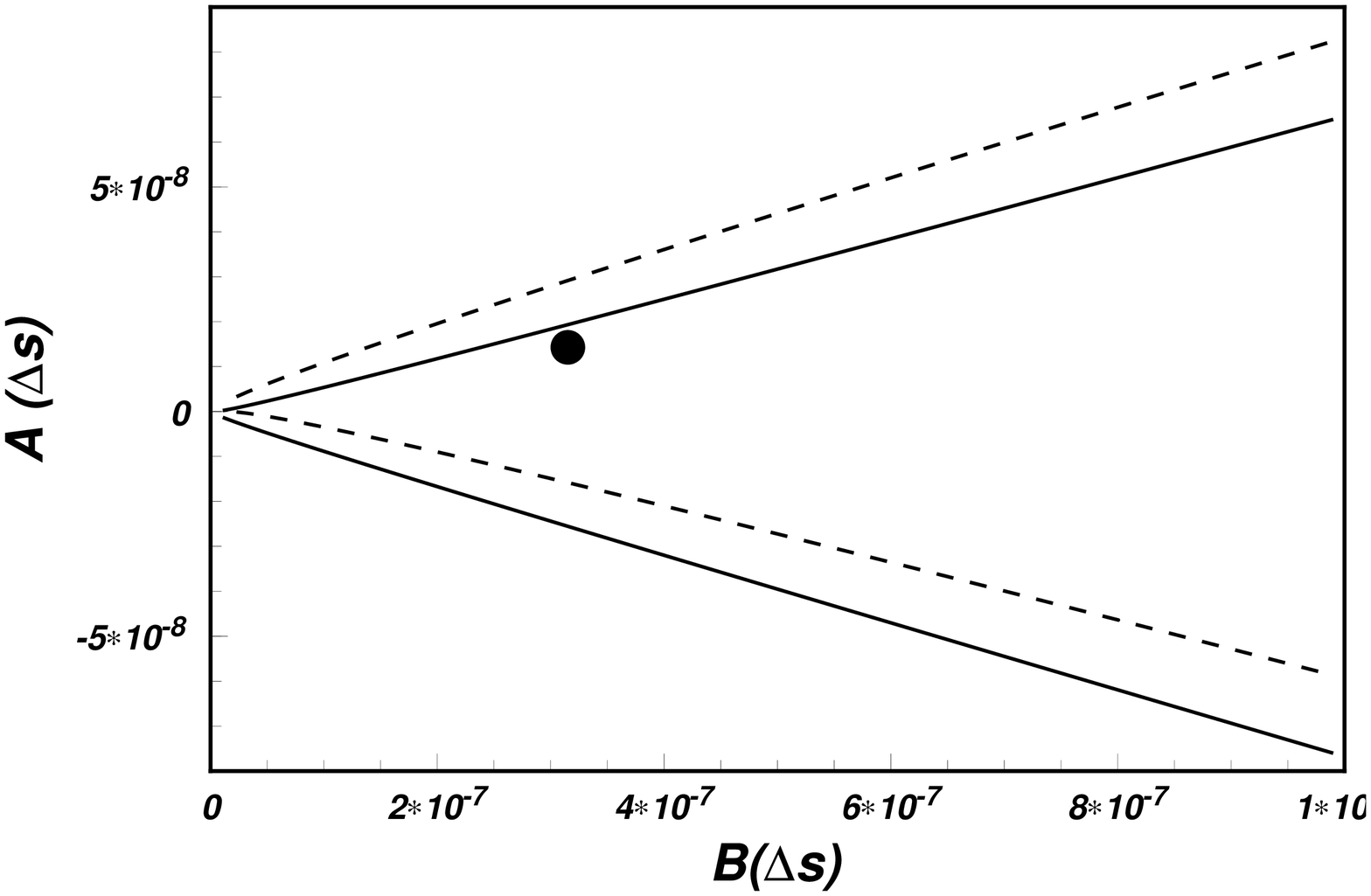}}
   \centerline{\parbox{11cm}{\caption{\label{liml}
The bounds for the partially integrated asymmetry
${\cal A}(\Delta s)$
  as a function of the partial dilepton branching ratio
   ${\cal B}(\Delta s)$.
The upper plot
is for the ``low-invariant-mass region"
and the lower one is for the 
``high-invariant-mass region" as defined in the text.
The boundary demarcated by the solid curves are for the
positive value  $C_7=0.3$, the  long-dashed curves correspond 
to the negative value
$C_7=-0.3$. The dotted line is the trivial bound
$|{\cal A}| < {\cal B}$. The solid dot denotes the SM value.
 }}}
\end{figure}

The upper and lower limits for the asymmetry as a function of
the branching fraction may be obtained analytically and are given by

\begin{equation} \label{bound}
{\cal A} (\Delta s)_\pm = \bar{R} \cos \theta_\pm
         \left( \alpha(\Delta s) \bar{R} \sin \theta_\pm + y \right) ,
\end{equation}
with
\begin{eqnarray}
\bar{R} &=& \sqrt{\frac{{\cal B} (\Delta s) - {\cal B}_{min} (\Delta
s)}
                     {A (\Delta s)}}  \\
\sin \theta_\pm &=& \frac{1}{4} \left[ -\frac{y}{\alpha(\Delta
s)\bar{R}}
\mp \sqrt{\frac{y^2}{\alpha^2(\Delta s)\bar{R}^2} + 8 } \, \right]  \\
y &=& \beta (\Delta s) - \frac{B (\Delta s) \alpha (\Delta s)}{2 A
(\Delta s)} .
\end{eqnarray}
These boundaries are shown in fig. \ref{liml} for the low-
and the high-invariant- mass region respectively. From these figures it
is obvious that the bound is in fact non-trivial: it lies in
any case far below the bound $|{\cal A}| < {\cal B}$.

\begin{figure}

   \vspace{-0.5cm}
   \epsfysize=11cm
   \centerline{\epsffile{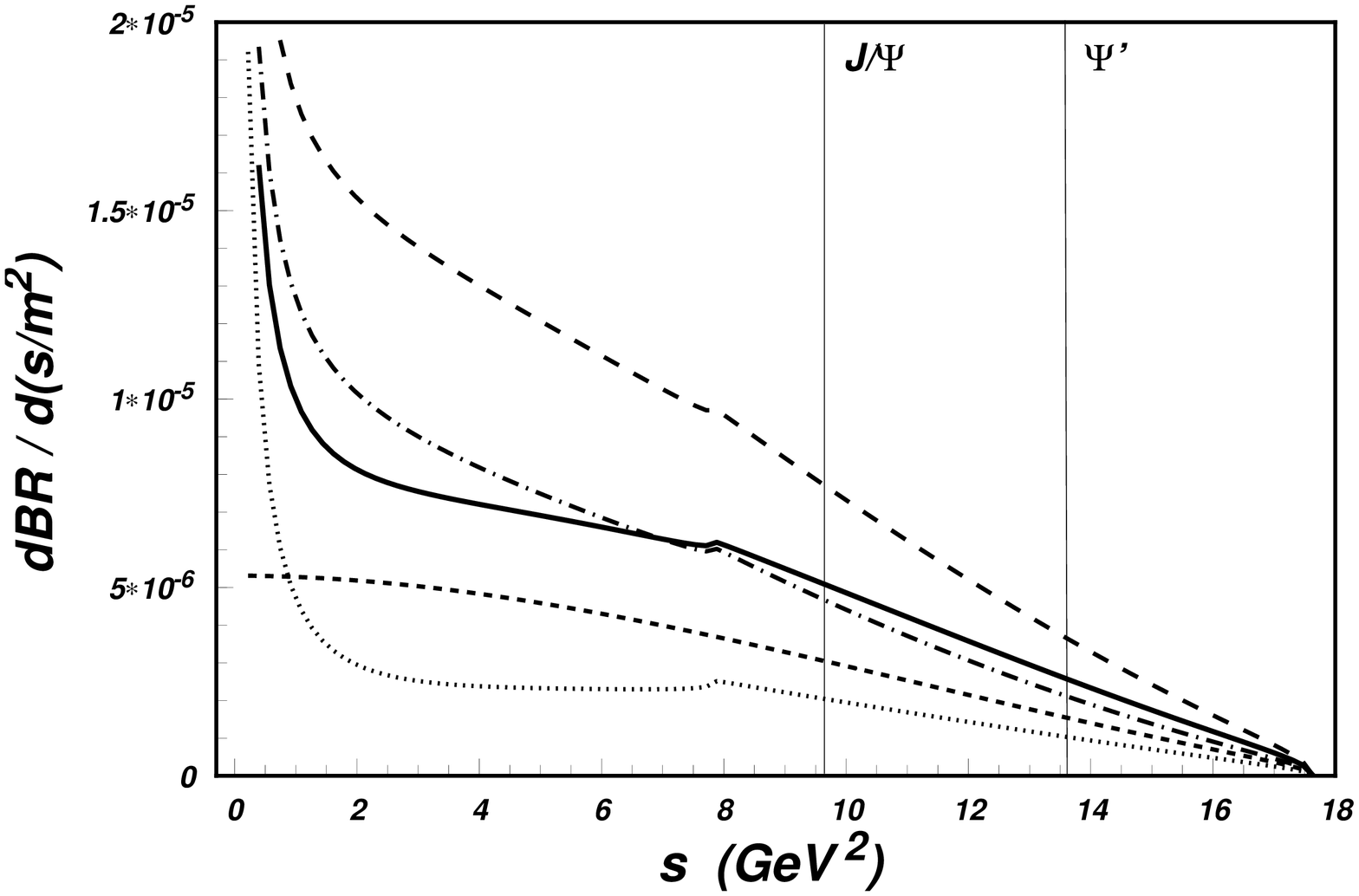}}
   \centerline{\parbox{11cm}{\caption{\label{fig5}
The dependence of the invariant-mass spectrum on the Wilson
coefficients.
Solid line: SM. Long-dashed line: $C_7 \to - C_7$, with other
coefficients retaining their SM values.
Short-dashed line: The contribution of $C_{10}$ only.
Dotted line: $C_{10} = 0$, with other coefficients retaining their SM values.
Dash-dotted line:  same as for the dotted one, but with $C_7 = -0.3$. 
The vertical lines indicate
the location of the $J/\Psi$ and $\Psi^\prime$ resonances. }}}
\end{figure}

If data become more precise one may in fact also expect a measurement
of the spectrum and the asymmetry as a function of $s$. The spectrum
itself is very sensitive to the values of the Wilson coefficients
and to the sign of $C_7$. In fig.~\ref{fig5} we plot the various
contributions to the spectrum, for positive and for negative $C_7$.

\begin{figure}

   \vspace{-0.5cm}
   \epsfysize=11cm
   \centerline{\epsffile{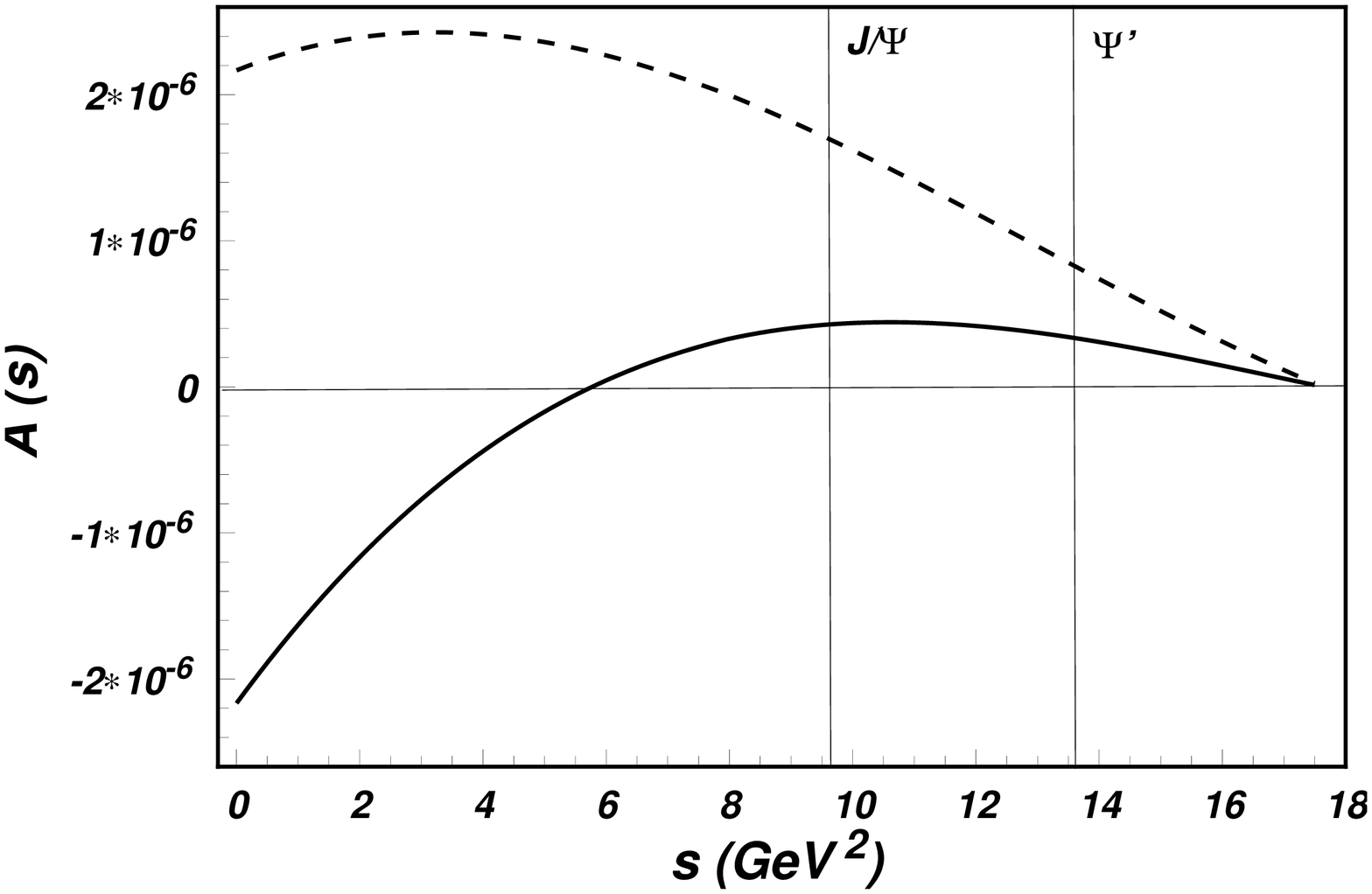}}
   \centerline{\parbox{11cm}{\caption{\label{fig6}
The dependence of the differential FB asymmetry on the Wilson
coefficients.
Solid line: SM. Long-dashed line: $C_7 \to - C_7$, with the other
parameters retaining their SM values.
The vertical lines indicate
the location of the $J/\Psi$ and $\Psi^\prime$ resonances.
   }}}
\end{figure}

In a similar way, it may become possible to measure also the
differential asymmetry

\begin{equation}
{\cal A} (s) = \int\limits_{-1}^0 dz \, \frac{d^2 {\cal B}}{d \hat{s}
d z}
- \int\limits_0^1 dz \, \frac{d^2 {\cal B}}{d \hat{s} d z} ,
\end{equation}
which is also sensitive to the sign of $C_7$. The various
contributions to ${\cal A} (s)$ are shown in fig.~\ref{fig6}.

%
%
\section{Model Predictions for the Wilson Coefficients}

Once the parameters $C_7$, $C_9$, and $C_{10}$ have
been extracted from experimental data, a direct comparison with
different theoretical predictions can be made. If a deviation
from the SM result is observed, new models  
accounting for this discrepancy, can be looked for. If no such
deviation is observed, the result can be used to set bounds on
new physics.

As already alluded to in the introduction, the operator basis
considered here is not the most general one. For example, in the
left-right symmetric (LR) models  the basis
(\ref{heff}) will have to be enlarged to
incorporate the extended $SU(2)_L \times SU(2)_R \times U(1)$ gauge
sector. The enlarged set contains the SM operators contained in
${\cal H}_{eff}$ above and another set of 10 operators obtained from
this by flipping the chirality structure of the fermion fields
$P_L \to P_R$, where $P_{L,R}= (1\pm \gamma_5)/2$, increasing the
number of independent Wilson coefficients to 20.

Since we are restricting the operator basis to exclude {\it ab initio}
LR-symmetric models, we give reasons why such extensions of the SM are
unlikely to make a significant contribution to rare $B$ decays.
In the minimal LR-symmetric models, the right-handed CKM matrix is
identical to, or the complex conjugate of, the left-handed CKM matrix.
However, in this case, there are rather strong lower limits on the mass
of $W_R$, $M_R  > 1.5 - 2.5 $ TeV, arising from the
condition that the short-distance contributions to the $K_L$--$K_S$ mass
difference not exceed the experimental value \cite{LRdmk}.
In view of this restriction
on $M_R$, it would be difficult to see tangible differences between the
SM and the minimal LR models in rare $B$ decays, much as
has been argued for the \bbar ~mixing ratio $x_s$ and $x_d$
\cite{Altfranz}. These qualitative anticipations have been borne out by
explicit calculations \cite{Jhewett,Rizzo94}.

The situation with the non-minimal LR models is more
involved \cite{Langacker}. While the bounds on $m_R$ from
$\Delta m_K$ can be evaded, the constraints from $\epsilon_K$ force
$m_R$ to be in excess of 30 TeV, in general \cite{LW89}.
So, we shall no longer
entertain the LR models, asserting that the present and
impending
constraints following from \brbgamaxs , $x_d$ and $\Delta m_K$ 
render the LR-symmetric-model effects in the FCNC
semileptonic decays \Bsell ~insignificantly small
and restrict the operator basis to
the one given in (\ref{heff}).
In this section, we consider as illustrative examples
two popular extensions of the SM (the MSSM and the
2HDM) and study how the predictions for
the Wilson coefficients $C_7$, $C_9$, and $C_{10}$ are altered in
these models, compared to the SM.
 
\begin{figure}

   \vspace{-0.5cm}
   \epsfysize=9cm
   \centerline{\epsffile{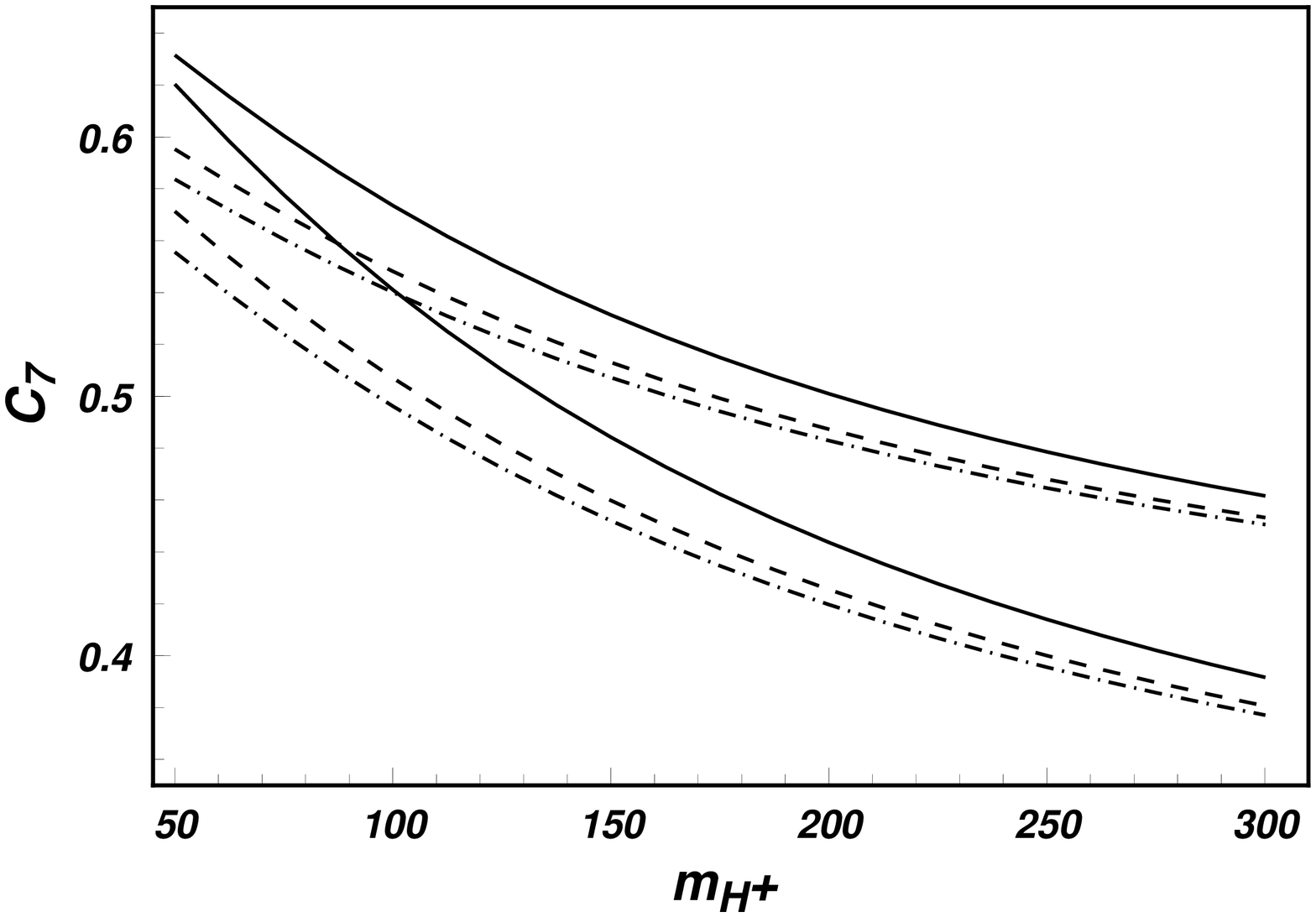}}
   \centerline{\parbox{11cm}{\caption{\label{higgs1}
The dependence of $C_7$ on the charged Higgs mass for
$v_2/v_1=1$ (solid line),
$v_2/v_1=2$ (dashed line),
$v_2/v_1=10$ (dot-dashed line).
The band between the two curves reflects the QCD uncertainty (see text).
   }}}
\end{figure}

Supersymmetric models have a new source of flavour-changing
neutral currents because of the quark-squark-gluino
vertex, which is, in general, not diagonal in generation
space \cite{fcnc}. If the generation mixings of such vertices
were arbitrarily large, they would lead to phenomenologically
unacceptable flavour violation in the $K^0$--${\bar{K}^0}$ system.
However, in the minimal supersymmetric model, flavour universality
of the supersymmetry-breaking terms is usually assumed to hold
at the grand-unification scale. In this case, a non-vanishing
(and calculable) generation mixing in the quark-squark-gluino vertex
is induced only by the effect of renormalization from the
grand-unification
scale to the weak scale. Given the present experimental limit on
the gluino mass, this effect leads only to very small contributions
to flavour-changing $b$decays \cite{bbmr} and it will be neglected
in our analysis. Gluino-mediated flavour-changing neutral currents
may however play an important role in non-minimal supersymmetric
models (see \cite{hag} and references therein).

\begin{figure}

   \vspace{-0.5cm}
   \epsfysize=9cm
   \centerline{\epsffile{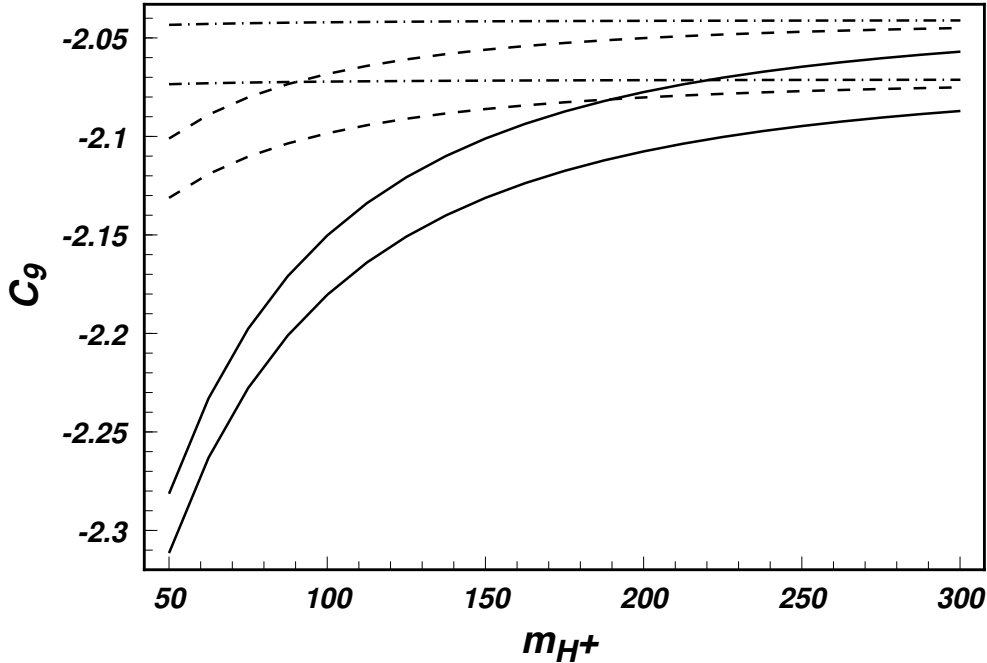}}
   \centerline{\parbox{11cm}{\caption{\label{higgs2}
The dependence of $C_9$ on the charged Higgs mass. The notation is
the same as in fig.~\protect{\ref{higgs1}}.
   }}}
\end{figure}

Once the gluino contributions (as well as the analogous ones from
neutralino exchange) are neglected, the flavour violation in the
model is completely specified by the familiar
CKM matrix. The one-loop supersymmetric corrections
to the Wilson coefficients
$C_7$, $C_9$, and $C_{10}$ are given by two classes of diagrams:
charged-Higgs exchange and chargino exchange. Their analytical
expressions can be found in ref. \cite{bbmr}, and will not be
repeated here.

The charged-Higgs contribution is specified by two input parameters:
the charged-Higgs mass ($m_{H^+}$) and the ratio of Higgs vacuum
expectation values ($v_2/v_1\equiv \tan \beta$). This
contribution is interesting
in itself, since it corresponds to a well-defined SM extension:
the 2DHM.
The interaction between the charged Higgs and quarks is given by:
\begin{equation}
{\cal{L}}=\frac{g}{\sqrt{2}m_W}H^+{\bar{u}}\left(
A_u M_uV\frac{1-\gamma_5}{2}+
A_d VM_d\frac{1+\gamma_5}{2}\right) d +\mbox{h.c.},
\end{equation}
where $M_{u,d}$ are the up and down quark mass matrices and $V$ is
the CKM matrix. If both up and down quarks get masses from the same
Higgs doublet (this case is usually referred to as Model I), then
\begin{equation}
A_u=-A_d=1/\tan \beta .
\end{equation}
In the supersymmetric case, two different Higgs doublets
couple separately to up and down quarks (Model II) and
\begin{equation}
A_u=1/A_d=1/\tan \beta .
\end{equation}

\begin{figure}

   \vspace{-0.5cm}
   \epsfysize=9cm
   \centerline{\epsffile{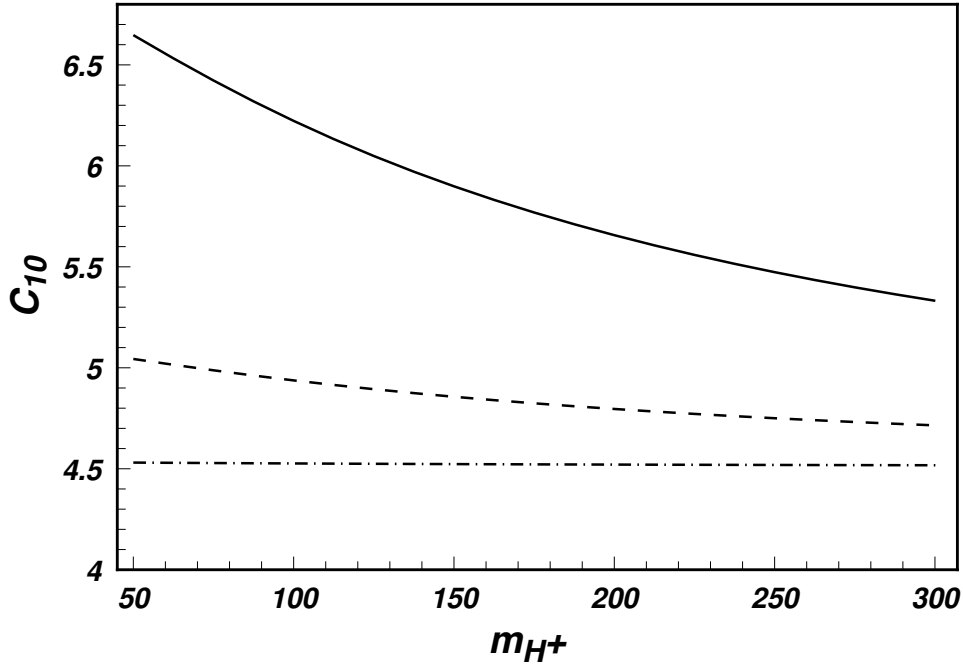}}
   \centerline{\parbox{11cm}{\caption{\label{higgs3}
The dependence of $C_{10}$ on the charged Higgs mass. The notation is
the same as in fig.~\protect{\ref{higgs1}}.
   }}}
\end{figure}

Let us first consider the case relevant to supersymmetry, i.e.
Model II.
Figures~\ref{higgs1}, \ref{higgs2} and \ref{higgs3} show the 
charged-Higgs effects on
$C_7$, $C_9$, and $C_{10}$ for $m_t=174$ GeV, at
leading order in QCD. For each value of $v_2/v_1$, two curves for
$\eta \equiv \alpha_s(M_Z)/ \alpha_s(\mu )$ equal to 0.68 and 0.41
are shown. The band within the two lines reflects the QCD
uncertainties,
since the two chosen values of $\eta$ correspond to the uncertainties
in
$\alpha_s(M_Z)$ ($=0.118 \pm 0.006$) \cite{Bethke} 
and $\mu$ ($m_b/2<\mu <2m_b$).
Recently the known part of the next-to-leading corrections has been
included in the analysis of $b\to s \gamma$ \cite{Martinelli}. They
have the
effect of reducing the scale dependence, and of decreasing considerably
the value of $C_7$ with respect to the leading-order calculation
shown in fig.~\ref{higgs1}. We expect that we can safely ignore
the next-to-leading corrections to $C_9$,
because the QCD running is much less important here than
in $C_7$.
Note that $C_{10}$
does not get renormalized under QCD.

The constraints imposed by supersymmetry on the scalar potential
require $\tan \beta >1$ and $m_{H^+}>m_W$.
Notice that, as $\tan \beta$ becomes large, $C_9$
and $C_{10}$ rapidly converge to the SM value (as $\tan^{-2}\beta$),
while the charged-Higgs
contribution to $C_7$ remains non-vanishing even in the limit
$\tan \beta \to \infty$.
In the case of a non-supersymmetric Model II, the experimental limit on
$BR(b\to s \gamma )$ rules out a large region of the 
$\tan \beta$--$m_H^{+}$
parameter space \cite{bsg,bur}. As a consequence,
no significant deviation from the SM values of $C_9$
and $C_{10}$ can be expected. In
supersymmetry, however, the chargino can largely compensate for
the charged-Higgs contribution to $C_7$ \cite{bg},
and more decisive
differences from the SM are foreseeable, as discussed below.

In the case of Model I, the expressions for $C_9$ and $C_{10}$
are still the same as in Model~II. However $C_7$
is modified and the constraint from $BR(b\to s \gamma )$ on
the parameter space is much weaker. For $\tan \beta <1$,
$C_7$ can become negative. For small $m_{H^+}$ and $\tan \beta$,
it is possible to reach values for which $C_7$ equals, in absolute
value, the SM prediction, but is opposite in sign. The region of
parameter space where this happens is barely allowed by the constraints
from $\epsilon_K$ and $B^0$--${\bar{B}^0}$ mixing \cite{burb}, but leads
to unacceptable
corrections to the $Z^0\to {\bar{b}}b$ width \cite{hol}
and is thus ruled out experimentally.

In addition to the diagrams with charged-Higgs exchange, the MSSM
leads also to chargino-mediated diagrams.
The chargino contribution is specified by six parameters.
Three of them enter the $2\times 2$ chargino mass matrix:
\begin{equation}
m_{\chi^+}=\pmatrix{M & m_W\sqrt{2}\sin \beta \cr
 m_W\sqrt{2}\cos \beta & \mu }.
\end{equation}
Following standard notations, we call $\tan \beta$ the ratio of
vacuum expectation values, the
same that appears also in the charged-Higgs sector, and
$M$, $\mu$ the gaugino and higgsino mass parameters,
subject to the constraint that the lightest chargino mass satisfies
the LEP bound, $m_\chi^+>45$ GeV. The squark masses
\begin{equation}
m^2_{{\tilde{q}}^2_\pm}={\widetilde{m}}^2+m_q^2\pm A{\widetilde{m}}m_q
\label{sq}
\end{equation}
contain two additional free parameters besides the known mass of
the corresponding quark $m_q$: a common supersymmetry-breaking mass
$\widetilde{m}$ and the coefficient $A$.
The parameter $A$ contains all the information (both from the
$\mu$-term and the trilinear term) of left-right
squark mixing and it is constrained
by the requirement that the
lightest stop is not produced at LEP, $m_{\tilde{t}}>45$
GeV \footnote{For
   a particular choice of the mixing between the two
   stop states, the light stop can be decoupled from the $Z^0$ and the
   LEP bound would not apply. We disregard here this possibility.}.
We also take into account
the CDF limit on squark masses \cite{cdf}, and impose $\widetilde{m}>126$
GeV. In eq.~(\ref{sq}) we have made the simplifying assumption that the
supersymmetry-breaking
left- and right-squark masses are equal.
The last parameter included in our analysis is a common mass $m_{\tilde{l}}$
for sleptons, all taken to be
degenerate in mass, with the constraint $m_{\tilde{l}}> 45$ GeV.
Therefore the version of the MSSM we are considering is defined in terms
of seven free parameters.

\begin{figure}[p]

   \vspace{-0.5cm}
   \epsfysize=9cm
   \centerline{\epsffile{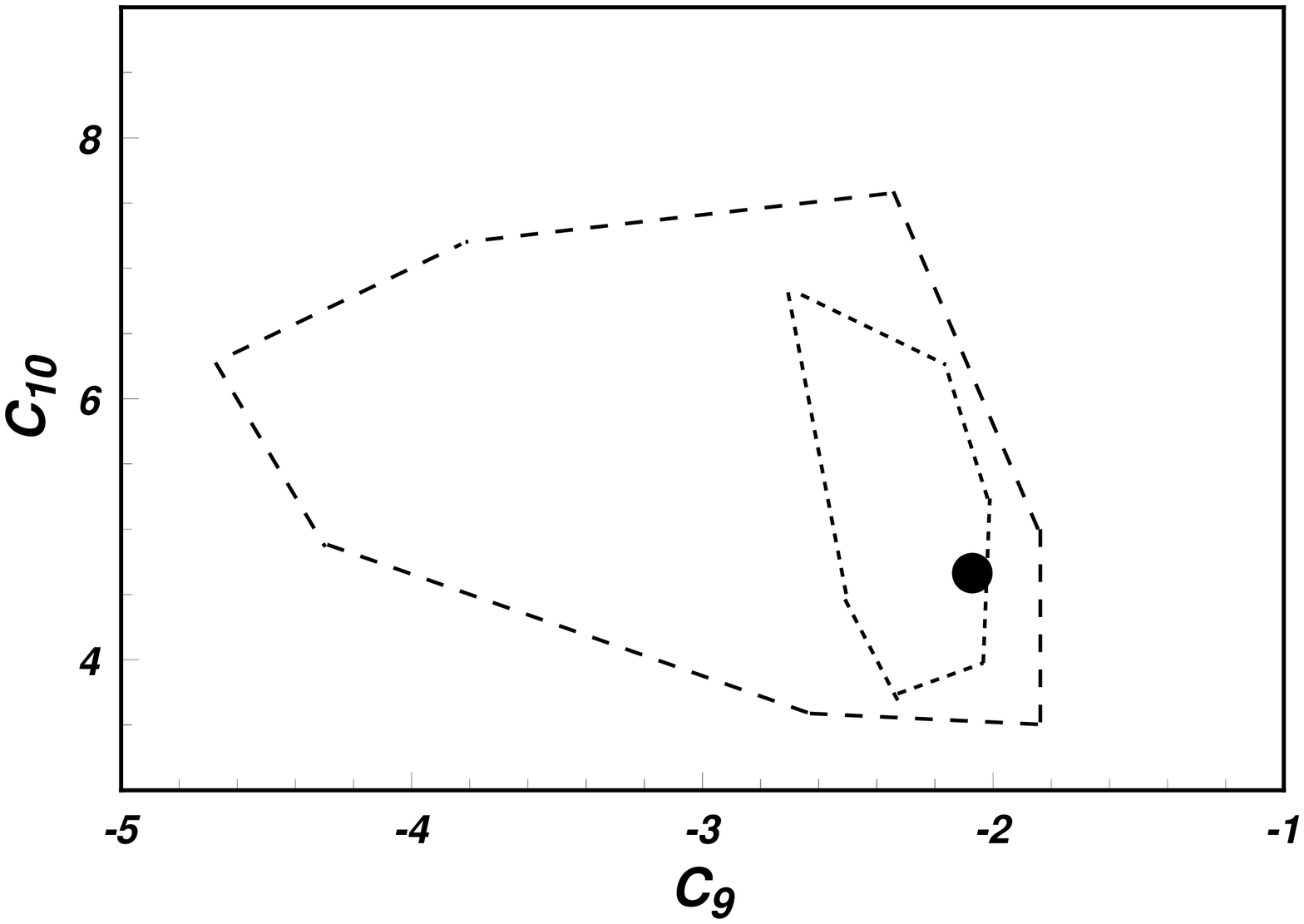}}
   \epsfysize=9cm
   \centerline{\epsffile{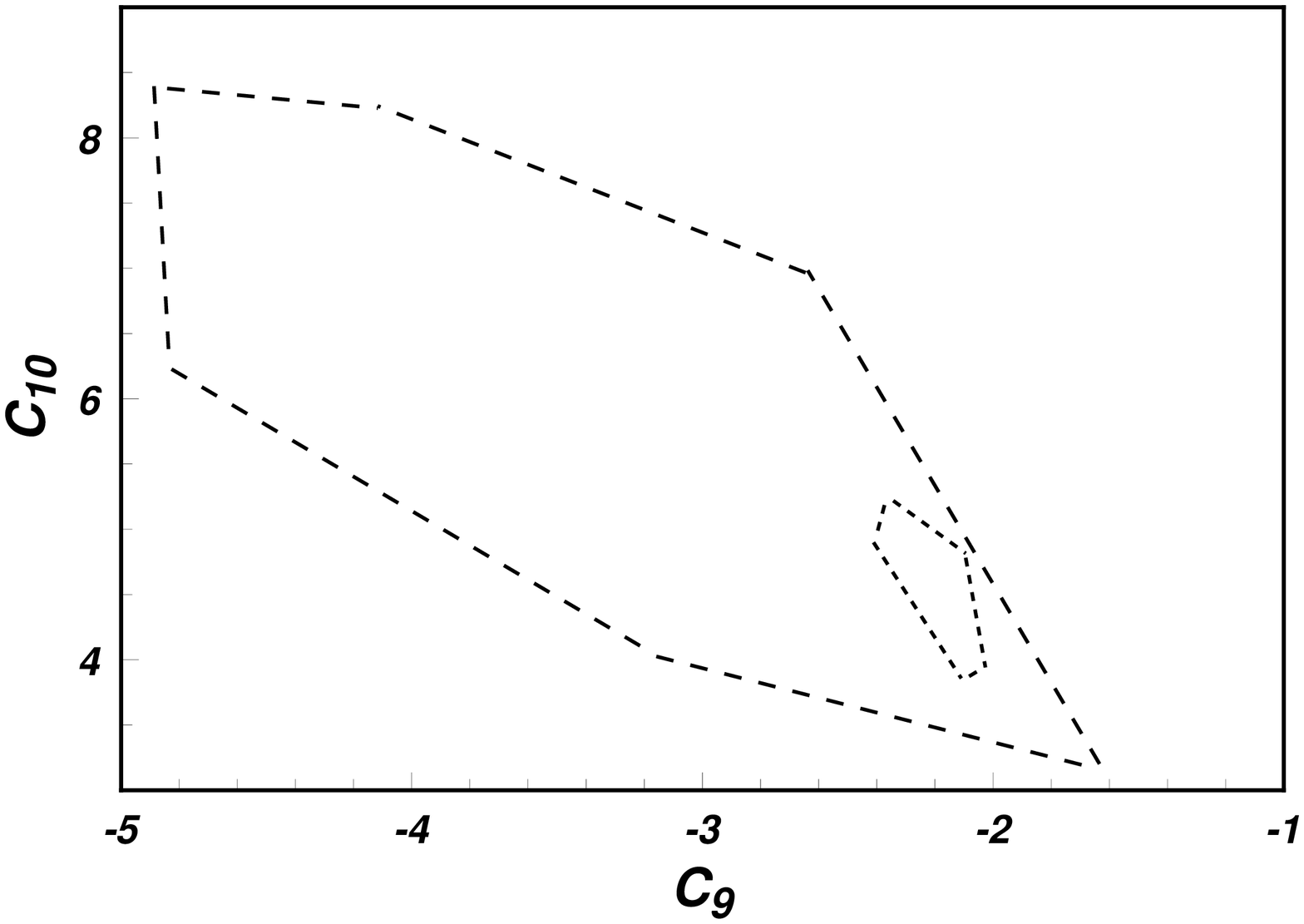}}
   \centerline{\parbox{11cm}{\caption{\label{susy1}
The region in the $C_9$-$C_{10}$ plane
obtained by varying the MSSM parameters.
The upper (lower) plot corresponds to solutions that satisfy
the $b\to s\gamma$ experimental constraint with positive (negative)
$C_7$ given in eq.(\protect{\ref{c7bound}}) and the present bounds 
($m_{H^+} > $ 80 GeV, $\tilde{m}_t, m_{\chi^+}, \tilde{m}_\ell > $ 45 GeV).
The smaller areas limited by the short-dashed line correspond to the
region of the MSSM parameter space that will survive an unsuccessful
search for supersymmetry at the Tevatron and LEP 200 ($m_{H^+}>150$ GeV,
$m_{\tilde{t}},~ m_{\chi^+},~m_{\tilde{l}}>100$ GeV).
   }}}
\end{figure}

We have computed the Wilson coefficients in the MSSM
and then varied the seven above-defined parameters in
the experimentally allowed region.
The results of our analysis are presented in fig.~\ref{susy1},
which shows the regions of the $C_9$--$C_{10}$ plane allowed by
possible choices of the MSSM parameters. The upper plot of 
fig.~\ref{susy1} corresponds
to parameters which give rise to positive (same sign as in the SM)
values of $C_7$, consistent with experimental results on $b\to
s\gamma$ ($0.19<C_7<0.32$), while the lower plot 
corresponds to
values of $C_7$ with opposite sign ($-0.32<C_7<-0.19$). We also
show how our results are affected
by an improvement in
the experimental limits on supersymmetric particle
masses, as can be expected from the Tevatron and LEP 200.
Fig.~\ref{susy1} also shows the $C_9-C_{10}$ regions allowed by
the MSSM if the further
constraints $m_{H^+}>150$ GeV, $m_{\tilde{t}},~ m_{\chi^+}
,~m_{\tilde{l}}>100$ GeV are imposed.
 
The regions shown in fig.~\ref{susy1} illustrate the
typical trend of the supersymmetric corrections. If supersymmetric
particles exist at low energies, we can expect larger values of $C_{10}$
and smaller (negative) values
of $C_9$ than those predicted by the SM.
This is the general feature, although the exact boundaries
of the allowed regions depend on the particular model-dependent
assumptions one prefers to use. For instance, the allowed regions
can be slightly expanded if one introduces more general stop mixings than
those considered here. On the other hand, if further constraints
are imposed on the model (such as particular boundary conditions at the
GUT scale, radiative symmetry-breaking, etc.), some of the
parameters which in our analysis were taken as independent
variables become related with each other, and the allowed region
may somewhat shrink. However, the most interesting feature of
supersymmetry is that solutions with negative values of $C_7$ are possible
and are still consistent with present data. Moreover, values of the 
other two coefficients $C_9$ and $C_{10}$ sufficiently different from 
the SM are allowed, leading to measurable differences in the decay rates 
and distributions of $B \to X_s \ell^+ \ell^-$ and $B_s \to \ell^+ \ell^-$.


\vspace*{1.2cm}
\noindent
\section{ Concluding Remarks}

The  branching ratios \brbksgam ~ and $ B \to X_s \gamma$ have provided  
first measurements of the strength of the effective magnetic moment
operator $\mb \bar{s}_L \sigma_{\mu \nu} b_R F^{\mu \nu }$.
The inferred value of the Wilson coefficient
 $\vert C_7(\mb) \vert = $0.19 -- 0.32 from present data is in
broad agreement with the SM estimate of the same. This, in turn, has
led to rather stringent bounds on the parameters of a number of
models. The comparison between the SM and experiment has become 
more precise with the measurement of the inclusive rate \brbgamaxs. 
There is an
overriding need to firm up theoretical estimates in the SM by
calculating the next-to-leading-order effects.

While the first step in the measurements of
rare $B$ decays has been made, we expect
this field to undergo a qualitative change as the
anticipated experimental facilities for $B$ physics take shape and
start producing $B$ hadrons in large quantities. We have in mind here
a steady consolidation of the CLEO and CDF data, with big strides
expected to be made at the threshold $B$ factories at SLAC and KEK and experimental
facilities with proton beams such as HERA-B and in particular the LHC.
Among others, we expect
measurements of the CKM-suppressed radiative decays
\brogam ~and \bomegam , and the FCNC semileptonic and leptonic decays
\Bsell ~ (and the corresponding exclusive decays 
$B \to (K, K^*) \ell^+ \ell^-$)
and $B_s^0 \to \ell^+ \ell^-$ at these facilities. Also, the
nature of rare $B$-decay measurements will evolve  in time from being
exploratory to becoming rather precise.
This has encouraged us to propose
undertaking a more ambitious
programme of extracting from data model-independent quantities -- very
much along the same lines as was carried out for the precision
electroweak analysis of the LEP, SLC and low-energy data.

The electroweak precision tests, at LEP and elsewhere, have concentrated on
the self-energies of the electroweak gauge bosons ($\gamma, Z^0$ and
$W^\pm$), with possible deviations from the SM expressed in terms of
a limited number of parameters \cite{Peskin,Altbarb}.
The main interest in rare $B$ decays is to measure the effective FCNC
vertices to test the SM precisely and  search for new physics.
We have argued how to parametrize these vertices through a
limited number of effective parameters, which govern the rates and
shapes (differential
distributions) in rare $B$ decays $B \to X_s \gamma$, \Bsell and 
$B_s \to \ell^+ \ell^-$. The search for
physics beyond the SM in these decays can be carried out in terms of
three effective parameters, which can then be interpreted in a large
class of models. The presence of non-SM physics may manifest itself by
distorting the differential distributions in \Bsell . Some possible
examples of such distortions have been worked out. While the analysis
presented here covers a large class of models, we have also presented
profiles of the Wilson coefficients in the best-motivated 
extensions of the SM, namely the MSSM. We have pointed out that there 
exist two distinct solutions corresponding 
to the negative and positive values for
$C_7(\mu)$, which cannot be distinguished from the data on
\brbgamaxs , but they give rise to very different distributions in the
decay \Bsell . Moreover, the coefficients of the 
operators $C_9(\mu)$ and $C_{10}(\mu)$ for the MSSM models may
attain values  sufficiently different from the corresponding SM values.
An encouraging result of our analysis is that
rare $B$ decays \bgamaxs ~and \Bsell ~have a discovery potential much
beyond direct searches at these facilities. Finally, we remark that
the purely leptonic decay modes $B_s^0 \to \ell^+ \ell^-$ can also be
used to further pin down the coefficient of the
operator ${\cal O}_{10}$ involving the axial-vector leptonic current, i.e.
$\vert C_{10}(\mu) \vert$, although this may require much larger $B$-hadron
statistics. Precision measurements in
rare $B$ decays must be pursued vigorously. 
\section*{Acknowledgements} 
We would like to thank Volodya Braun, Christoph Greub,
Matthias Jamin, Guido
Martinelli and Mikolaj Misiak for helpful discussions and correspondence.



\appendix
\section*{Appendix: The effective Hamiltonian for $B \to X_s \gamma$ and
$B \to X_s \ell^+ \ell^-$}
The effective Hamiltonian is
\begin{equation}
H_{eff} = - \frac{4 G_F}{\sqrt{2}} V_{ts}^* V_{tb}
        \sum_{i=1}^{10} C_i (\mu) {\cal O}_i (\mu) ,
\end{equation}
where the operator basis is chosen to be
\begin{eqnarray}
{\cal O}_1 &=& (\bar{s}_{L \alpha} \gamma_\mu b_{L \alpha})
               (\bar{c}_{L \beta} \gamma_\mu c_{L \beta})     \\
{\cal O}_2 &=& (\bar{s}_{L \alpha} \gamma_\mu b_{L \beta})
               (\bar{c}_{L \beta} \gamma_\mu c_{L \alpha})     \\
{\cal O}_3 &=& (\bar{s}_{L \alpha} \gamma_\mu b_{L \alpha})
               \sum_{q=u,d,s,c,b}
               (\bar{q}_{L \beta} \gamma_\mu q_{L \beta})     \\
{\cal O}_4 &=& (\bar{s}_{L \alpha} \gamma_\mu b_{L \beta})
                \sum_{q=u,d,s,c,b}
               (\bar{q}_{L \beta} \gamma_\mu q_{L \alpha})     \\
{\cal O}_5 &=& (\bar{s}_{L \alpha} \gamma_\mu b_{L \alpha})
               \sum_{q=u,d,s,c,b}
               (\bar{q}_{R \beta} \gamma_\mu q_{R \beta})     \\
{\cal O}_6 &=& (\bar{s}_{L \alpha} \gamma_\mu b_{L \beta})
                \sum_{q=u,d,s,c,b}
               (\bar{q}_{R \beta} \gamma_\mu q_{R \alpha})     \\
{\cal O}_7 &=& \frac{e}{16 \pi^2} m_b
               (\bar{s}_{L \alpha} \sigma_{\mu \nu} b_{R \alpha})
                F^{\mu \nu}                                    \\
{\cal O}_7^\prime &=& \frac{e}{16 \pi^2} m_s
               (\bar{s}_{R \alpha} \sigma_{\mu \nu} b_{L \alpha})
                F^{\mu \nu}                                     \\
{\cal O}_8 &=& \frac{g}{16 \pi^2} m_b
(\bar{s}_{L \alpha} T_{\alpha \beta}^a \sigma_{\mu \nu} b_{R \alpha})
                G^{a \mu \nu}                                    \\
{\cal O}_8^\prime &=& \frac{g}{16 \pi^2} m_s
(\bar{s}_{R \alpha} T_{\alpha \beta}^a \sigma_{\mu \nu} b_{L \alpha})
                G^{a \mu \nu}                                    \\
{\cal O}_9 &=& (\bar{s}_{L \alpha} \gamma_\mu b_{L \alpha})
               (\bar{\ell} \gamma_\mu \ell)                      \\
{\cal O}_{10} &=& (\bar{s}_{L \alpha} \gamma_\mu b_{L \alpha})
               (\bar{\ell} \gamma_\mu \gamma_5 \ell)             \\
\end{eqnarray}
where
\begin{equation}
q_L = \frac{1-\gamma_5}{2} q  \mbox{ and }
q_R = \frac{1+\gamma_5}{2} q .
\end{equation}

The Wilson coefficients of the operators are given by the renormalization
group evolution
\begin{equation} \label{RGE}
\left[\mu \frac{\partial}{\partial \mu}
+ \beta(g) \frac{\partial}{\partial g} \right]
C_i \left(\frac{M^2_W}{\mu^2},g \right)
= \hat{\gamma}_{ji} (g) C_j \left(\frac{M^2_W}{\mu^2},g \right) .
\end{equation}
To leading logarithmic level, the  QCD beta function $\beta(g)$ is
given by
\begin{equation}  \label{bet}
\beta(g) = -\beta_0 \frac{g^3}{16 \pi^2} \,\, \mbox{ with } \,\,
\beta_0=11-\frac{2}{3}f
\end{equation}
and $\hat{\gamma}(g)$ is the anomalous dimension matrix, which is, to leading
logarithmic acuracy, given by
\begin{equation}  \label{gam}
\hat{\gamma}(g) = -\gamma_0 \frac{g^2}{16 \pi^2}
\end{equation}
Here $\gamma_0$ is a $10 \times 10$ matrix given by 
{\footnotesize
\begin{eqnarray*}
&& \gamma_0 =\\
&& \left[ \begin{array}{cccccccccc}
\vspace{0.2cm}
    -2        &         6        &        0        &         0     &
     0        &         0        &        0        &
     3        &  -\frac{16}{3}   &        0       \\
\vspace{0.2cm}
     6        &        -2        &-\frac{2}{9}     &  \frac{2}{3}  &
 -\frac{2}{9} &  \frac{2}{3}     & \frac{416}{81}  &
\frac{70}{27} &  -\frac{16}{9}   &        0       \\
\vspace{0.2cm}
     0        &         0        &-\frac{22}{9}    & \frac{22}{3}  &
 -\frac{4}{9} &  \frac{4}{3}     &-\frac{464}{81}  &
\frac{140}{27}+ 3 f
&-\frac{16}{3}\left(u-\frac{d}{2}-\frac{1}{3}\right)
& 0  \\
\vspace{0.2cm}
     0        &         0        &  6-\frac{2}{9}f &-2+\frac{2}{3}f&
-\frac{2}{9}f &  \frac{2}{3}f    &\frac{416}{81}u-\frac{232}{81}d  &
6+\frac{70}{27}f &-\frac{16}{3}\left(u-\frac{d}{2}- 3 \right) &   0   \\
\vspace{0.2cm}
     0        &         0        &          0     &      0     &
     2        &        -6        &  \frac{32}{9}  &
-\frac{14}{3}- 3 f    & -\frac{16}{3}\left(u-\frac{d}{2}\right) &
     0     \\
\vspace{0.2cm}
     0        &         0        &-\frac{2}{9}f   & \frac{2}{3}f   &
-\frac{2}{9}f &-16+\frac{2}{3}f  &-\frac{448}{81}u+\frac{200}{81}d &
-4-\frac{119}{27}f & -\frac{16}{9}\left(u-\frac{d}{2}\right)&   0    \\
\vspace{0.2cm}
     0        &         0        &          0     &         0     &
     0        &         0        &   \frac{32}{3} &
     0        &         0        &          0       \\
\vspace{0.2cm}
     0        &         0        &          0     &          0    &
     0        &         0        & -\frac{32}{9}  &
 \frac{28}{3} &         0        &          0      \\
\vspace{0.2cm}
     0        &         0        &          0     &          0    &
     0        &         0        &          0     &
     0        &         0        &          0      \\
\vspace{0.2cm}
     0        &         0        &          0     &          0    &
     0        &         0        &          0     &
     0        &         0        &          0
\end{array} \right]
\end{eqnarray*}}
Using as initial condition $C_j (M_W) = 0$ for
$j = 1, 3, \cdots 6$ we obtain, for the solution of the renormalization
group flow
\begin{eqnarray}
C_1 (\mu ) &=& \frac{1}{2} C_2 (M_W)
               \left( \eta^{6/23} -  \eta^{-12/23} \right)
\\
C_2 (\mu ) &=& \frac{1}{2} C_2 (M_W)
               \left( \eta^{6/23} +  \eta^{-12/23} \right)
\\
C_3 (\mu ) &=& C_2 (M_W) \left(-0.0112 \eta^{0.8994}
                         + \frac{1}{6} \eta^{-12/23}
                         - 0.1403 \eta^{-0.4230}
                         + 0.0054 \eta^{0.1456} \right.
\\ \nonumber
           && \qquad \qquad  \left. \vphantom{\frac{1}{6}}
                         - 0.0714 \eta^{6/23}
                         + 0.0509 \eta^{0.4086} \right)
\\
C_4 (\mu) &=& C_2 (M_W) \left(0.0156 \eta^{0.8994}
                         - \frac{1}{6} \eta^{-12/23}
                         + 0.1214 \eta^{-0.4230}
                         + 0.0026 \eta^{0.1456} \right.
\\ \nonumber
           && \qquad \qquad  \left. \vphantom{\frac{1}{6}}
                         - 0.0714 \eta^{6/23}
                         + 0.0984 \eta^{0.4086} \right)
\\
C_5 (\mu) &=& C_2 (M_W) \left(-0.0025 \eta^{-0.8994}
                         + 0.0117 \eta^{-0.4230}
                         + 0.0304 \eta^{0.1456}
                         - 0.0397 \eta^{0.4086} \right)
\\
C_6 (\mu) &=& C_2 (M_W) \left(-0.0462 \eta^{-0.8994}
                         + 0.0239 \eta^{-0.4230}
                         - 0.0112 \eta^{0.1456}
                         + 0.0335 \eta^{0.4086} \right)
\\
C_7 (\mu) &=& C_7 (M_W) \eta^{16/23} +
              C_8 (M_W) \frac{8}{3}
                        \left( \eta^{14/23} - \eta^{16/23} \right)
\\ \nonumber
          && + C_2 (M_W) \left(- 0.0185 \eta^{-0.8994}
                               - 0.0714 \eta^{-12/23}
                               - 0.0380 \eta^{-0.4230}
                               - 0.0057 \eta^{0.1456} \right.
\\ \nonumber
           && \qquad \qquad  \left.
                               - 0.4286 \eta^{6/23}
                               - 0.6494 \eta^{0.4086}
                               + 2.2996 \eta^{14/23}
                               - 1.0880 \eta^{16/23} \right)
\\
C_8 (\mu) &=& C_8 (M_W) \eta^{14/23}
\\ \nonumber
          && + C_2 (M_W) \left(- 0.0571 \eta^{-0.8994}
                               + 0.0873 \eta^{-0.4230}
                               + 0.0209 \eta^{0.1456} \right.
\\ \nonumber
           && \qquad \qquad  \left.
                               - 0.9135 \eta^{0.4086}
                               + 0.8623 \eta^{14/23} \right)
\\
C_9 (\mu) &=& C_9 (M_W) + C_2 (M_W) \left( \frac{37}{33}
                         - 0.0193 \eta^{-0.8994}
                         +  \frac{74}{999} \eta^{-12/23}
                         - 0.2608 \eta^{-0.4230} \right.
\\ \nonumber
           && \qquad \qquad  \left.
                         - 0.0183 \eta^{0.1456}
                         - \frac{10}{9} \eta^{6/23}
                         + 0.2143 \eta^{0.4086}\right)
\\
C_{10} (\mu ) &=& C_{10} (M_W) ,
\end{eqnarray}
where
$$
\eta = \frac{\alpha_s (M_W)}{\alpha_s (m_b)}
$$
and we use $C_2 (M_W) = -1$.

In terms of this effective Hamiltonian, the amplitude for 
$b \to s \ell^+ \ell^-$ becomes
\begin{eqnarray}
M&=& 4 {\sqrt 2} G_F {\alpha \over 4 \pi} (V_{ts}^* V_{tb})
 \left\{ C_{9\, eff} {\sLbar \gmu b_L} { \VV} 
   + C_{10} {\sLbar \gmu b_L} { \AA}  \right. \nonumber\\
& & \left. - C_7 \sbar i \sig {q^\nu \over q^2} (m_s b_L+ m_b b_R)  \VV \right\},
\label{bsllamp}
\end{eqnarray}
where  $q^\mu=p_1^{\mu}+p_2^{\mu}$ is the momentum transferred 
to the leptons. The effective coefficient $C_{9\,eff}$ contains the 
contributions from the one-loop matrix elements of ${\cal O}_1, ... ,
{\cal O}_6$ and is given by (\ref{c9eff}). This definition involves the 
one-loop function $g$, which is given by
\begin{eqnarray}
\mbox{ Re } g(z,s) &=& - \frac{4}{9} \ln z^2 + \frac{8}{27} 
                     + \frac{16z^2}{9s} \\ \nonumber  
&& - \frac{2}{9} 
\sqrt{1 - \frac{4 z^2}{s}} \left( 2 + \frac{4 z^2}{s} \right) 
\ln \left| \frac{1 + \sqrt{1 - \frac{4 z^2}{s}}}
                {1 - \sqrt{1 - \frac{4 z^2}{s}}} \right|
\,\, \mbox{ for } s > 4z^2   \\ 
\mbox{ Re } g(z,s) &=& - \frac{4}{9} \ln z^2 + \frac{8}{27} 
                     + \frac{16z^2}{9s} \\ \nonumber  
&& - \frac{2}{9} 
\sqrt{1 - \frac{4 z^2}{s}} \left( 2 + \frac{4 z^2}{s} \right) 
\mbox{ atan } \left( \frac{1} {\sqrt{\frac{4 z^2}{s} - 1}} \right)
\,\, \mbox{ for } s < 4z^2   \\
\mbox{ Im } g(z,s) &=& - \frac{2 \pi}{9} \sqrt{1 - \frac{4 z^2}{s}} 
                         \left( 2 + \frac{4 z^2}{s} \right) 
                         \Theta (s - 4z^2)  .
\end{eqnarray}
\end{document}